\documentstyle[epsbox,twocolumn]{mn}

\newcommand{\beq}{\begin{equation}}
\newcommand{\eeq}{\end{equation}}
\newcommand{\bea}{\begin{eqnarray}}
\newcommand{\eea}{\end{eqnarray}}
\title{Statistics of Merging Peaks of
Random Gaussian Fluctuations: Skeleton Tree Formalism}
\author[Hitoshi HANAMI]
{Hitoshi HANAMI \\
Physics Section, 
Faculty of Humanities and Social Sciences, 
Iwate University, Morioka 020 JAPAN}
\date{Accepted 1999 August ??
      Received 1999 July 14;
      in original form 1999 July 11}
\pagerange{\pageref{firstpage}--\pageref{lastpage}}
\pubyear{1999}

\begin{document}

\maketitle

\begin{abstract}
The cosmological bound objects were considered to form
from the local maxima of cosmological density fluctuations;
often assumed to be Gaussian random fields.
In order to study the statistics 
of the objects with hierarchical merging,  
we propose {\it the skeleton tree formalism},  
which can analytically
distinguish the episodic merging and the continuous accretion
in the mass growth processes.  The distinction was not 
clear in extended Press-Schechter (PS) formalism. 
The skeleton tree formalism 
is a natural extension of the peak theory which is an 
alternative formalism for the statistics of the bound objects.    
The fluctuation field smoothing with Gaussian filter produces 
the landscape with adding the extra-dimension of the filter resolution 
scale to the spatial coordinate of the original fluctuation.
In the landscape,
some smoothing peaks are nesting into the neighboring peaks
at a type of critical points called {\it sloping saddles}  
appears, which can be interpreted as 
merging events of the objects in the context of the
hierarchical structure formation.  The topological properties 
of the landscape can be abstracted in { \it skeleton trees},
which consist of line process of the smoothing peaks and
the point process of the sloping saddles.  
According to this abstract topological picture,
in this paper, we present the concept and the basic results
of {\it the skeleton tree formalism}  
to describe
(1) the distinction between the accretion and the merger in the
hierarchical structure formation from various initial 
random Gaussian fields;   
(2) the instantaneous number density of the sloping saddles which
gives the instantaneous scale function of the objects with the
destruction and reformation in the mergers;  
(3) the rates of the destruction, the reformation, and
the relative accretion growth;  
(4) the self-consistency of the formalism 
for the statistics of the mass growth processes of the
objects; 
(5) the mean growth history of the objects at the fixed mass.   
\end{abstract}

\begin{keywords}
galaxies:clustering -- galaxies:formation -- cosmology:theory -- dark matter
\end{keywords}

\section{INTRODUCTION} 
 
Hierarchical clustering scenario, including the cold dark matter
(CDM) model, may be the most established one for 
reconstructing various observational properties in the 
cosmological structure from the galaxies to the clusters of 
galaxies.
Press $ \& $ Schechter (1974) firstly proposed an analytical formalism
which derives the number density of bound virialized objects of the 
mass at any given epoch, with the assumption that the primordial 
density fluctuations is random Gaussian field. The mass function
predicted by the PS theory shows reasonably
the good agreement with N-body simulations even if it has more low
mass objects (e.g. Lacey $\&$ Cole 1994).   
To reconstruct the observational properties in  
theoretical galaxy formation scenario,  
there are also approaches which study the history of the mass growth for 
bound objects and the characteristic times (e.g. Lacey $ \& $ Silk 1991; 
Kauffmann, White  $ \& $ Guiderdoni 1993; Cole et al. 1994).
Most of them were based on the extended PS formalism, which was
proposed by Bower (1991) and Bond et al. (1991).  It
can derive the number density of objects of a certain mass at a 
given time subject to a larger object at a later time.  Using the 
formalism, Lacey $ \& $ Cole (1993; LC) calculated the ``merger'' rate.  

The PS formalism, however, has a limitation for 
describing the history of the mass growth about the individual objects.  
The ``merging process'' described with the PS approach in LC, cannot be 
interpreted as the same meaning of the merger in astronomical sense, 
in which the objects lose their identity.  
In the mass growth history for the astronomical objects, 
the continuous accretion onto a bound object  
without the loss of identity has different meanings from 
the mass accumulation with the loss of the identity in the major merger.  
The formalism based with the
PS approach cannot imply the distinction between ``tiny'' and ``notable'' 
captures.  For solving this problem,
Manrique $ \& $ Salvador-Sole (1995, 1996) proposed a 
formalism named ``ConflUent System of Peak trajectories'' (CUSP)
formalism as an extension of adaptive windowing
by Appel and Jones (1990).  They can be categorized into
a type of ``peak'' theory (Doroshkevch 1970; Adler 1981;
Bardeen et al. 1986, hereafter BBKS), 
which can count the number of  
the peaks of the density related to 
the collapsing threshold, applying a
low-pass filter of the bound object scale to the fluctuation field.  
The CUSP formalism, unfortunately, needs the iterative 
calculation to estimate the destruction
and the reformation rates with mergers. This point 
becomes a disadvantage 
when we try to apply this scheme to the semi-analytical
studies for the galaxy formation, including
the mass accumulation history of bound 
objects with Monte Carlo method as shown in LC.  

In order to give analytical description 
for the destruction and the reformation rate in the merger,
we developed a new approach called 
{\it the skeleton tree formalism}, using the
topological characteristics in the smoothing of the random field
with Gaussian filter.
With the appearance of {\it
sloping saddles} in the landscape of the smoothed field,
we can pick up the merging events, 
and distinguish the merger and the accretion.
The topological feature in the landscape 
with the sloping saddles can be 
extremely reduced into {\it skeleton trees}.

In this paper, we will focus on the concept and the
basic description of {\it the skeleton tree formalism}.  The outline of the
paper is as follows. In Section 2, 
in order to distinguish the
accretion and the merging in our
context, we sketch the topological characteristics in the landscape
of the smoothed field with critical points; the peaks and the sloping saddles,
define the merger events with appearance 
of the sloping saddles,  and reconstruct the growth history of the objects
with {\it skeleton tree} picture.
In Sections 3, 4, and 5, we formulate the
mathematical description of the constraints, the
probability distribution functions, and the scale functions
for the critical points.   In Section 6, we
shows the results of the evolution rates with the
accretion and the merging obtained from {\it the skeleton tree formalism}
with its consistency.
Finally, we present our conclusion in Section 7.  We have relegated
the details of the derivations to five appendices.  

\section{Identification of Accretion and Merging}

\subsection{Hierarchical Evolution from Fluctuations and Filtering Process} 

We will express the density fluctuation field 
as the functions of the comoving 
spatial coordinate $ \mbox{\bf r} $ and $ \mbox{\bf k} $ ; 
\bea 
 \delta(\mbox{\bf r})= \int d^3\mbox{\bf k} 
e^{i {\mbox{\bf  k r}} } \delta ({\bf k}) \; .
\eea 
Interested collapse objects of a comoving scale $ R $ 
can be identified as peaks greater than a threshold, whose 
fields are smoothed with a low-pass filter of the resolution scale $ R $.  
The fluctuation, smoothed with the selection function
$ S(\mbox{\bf r};R) $,  
can be expressed as 
\beq 
F(\mbox{\bf r};R) = \frac{\int d^3\mbox{\bf  r}_0 
S(\mbox{\bf r}_0;R) 
\delta (\mbox{\bf  r}- \mbox{\bf  r}_0)} 
{ \int d^3\mbox{\bf  r}_0 S(\mbox{\bf  r}_0;R) } \; . 
\eeq 
This Fourier transform is represented with the window function which 
is the Fourier transforms of the selection function; 
\beq
F(\mbox{\bf k};R) = \delta({\mbox{\bf k}}) 
W(\mbox{\bf k};R) \; .  
\eeq
In our interest cases, the window function works as a low-pass filter.

We shall restrict ourselves to isotropic homogeneous Gaussian random
fields with zero mean as descriptions of the initial fluctuations.  
For the field, the power
spectrum is then only a function of $ k = \vert \mbox{\bf k} \vert $;
$\vert \delta_k \vert^2 = \vert \delta ({\bf k}) \vert$.  
The fluctuation spectrum filtered with the scale $ R$ is 
\beq 
P(k;R)=\vert \delta_k \vert^2 W(k;R)^2 \; .
\eeq 
In this paper, 
we take a normalized isotropic Gaussian filter:  
\beq
W(\mbox{\bf k};R) = W(k;R) \equiv e^{ -k^2 R^2/2 } \; . 
\eeq
In the next subsection, we will
discuss the reason related to its unique property for the
smoothing.  

In the linear theory of gravitational instability
for the structure formation, 
the amplitude of the field in the overdensity area 
firstly grows in proportion to $ D(t) $, 
where $ D(t) $ is the linear growth factor.
According to BBKS, a 
bound object collapses from the area of a comoving scale $ R $
when the density of a peak in the 
fluctuation smoothed over the resolution scale
$ R $ exceeds above a fixed threshold
$ \delta_{c;0} $.  
Instead of viewing the peaks to be growing in density 
amplitude relative to the fixed threshold $ \delta_{c;0} $,  
we can interpret that the threshold level $ \delta_c $ is decreasing 
as $ \delta_{c;0} D(t_{0}) D(t)^{-1} $ with fixing 
the initial fluctuation field $ F(\mbox{\bf r};R) $, where 
$\delta_{c;0}$ was determined from the threshold at the
present time $t_0$.  
In this paper, we take Einstein-de Sitter
model: $ \Omega_0 =1, \lambda_0 =0 $, in which the relative threshold level
$ \delta_c / \delta_{c;0}= D(t_{0}) D(t)^{-1} = (1+z) $, where $ z $
is the redshift at the time $ t $.

For standard initial fluctuations like CDM models, in general, 
the rms of the smoothed field $ < F ({\bf r}; R) ^2 > $ is  decreasing
as increasing $ R $.   For such a fluctuation,
as decreasing the collapse threshold with time evolution, 
we can pick the collapse objects in the larger scale, which
gives a reasonable sketch of hierarchical clustering picture.
We will relate the filtering process and 
the hierarchical clustering description in the next.

\subsection{Landscape of the Fluctuation Field in Position 
and Resolution Space}  

Consider the random field in one dimensional (1-D) positional 
coordinate of $ x $ is smoothed with a low-pass filter of
a resolution scale $ R $.  This field
is reproduced as a landscape which extends in two dimensional (2-D)
extended space of $ (x, R) $.
A smoothing peak with a low-pass filter make a ridge  
which is running along the direction of increasing
$ R $ in the landscape.
The threshold 
level of the collapse can be interpreted as an ocean surface which makes a 
``shoreline'' and some ``lakes'' and ``islands'' in the landscape.
A ``cape'' on the shoreline of $ \delta_c $, 
can be identified as a bound object at $ \delta_c $.   
Then,  we can count the number of the bound objects of the scale $ R $ 
to pick up the capes at the $ R $ in the landscape.  
As the level of the ocean surface of $ \delta_c $ 
is decreasing with the evolution of the universe, 
the shoreline moves to the offing.  It means that the bound objects 
grow their scale continuously.  

However, if ``islands'' or ``lakes'' appear in the landscape,
they confuse the identification of ``capes'' as bound objects.
As an example, let consider 
an island in the offing of a cape on a shoreline. In this case, 
we have also another offing side cape of the island; there are two 
capes which can be counted as the bound objects of ``clouds''.  
In these situations, the cloud of the island cape  
contains the smaller cloud of the cape on the shore.   
These problems appear  
when we take general filters except the Gaussian filter.  
Bond et al. (1991) had shown the filter dependency of 
the landscape ( see their Figure 1 and 2 ), in which they 
take sharp $k$-space filter and Gaussian filter.
In the case of the sharp $k$-space filter, 
the ridges cannot decrease monotonically as 
the resolution scale $ R $ increases.  
On the other hand,  in the case of the Gaussian filter, the ridge decreases 
monotonically as the resolution scale $ R $ increases.  
They represent about this feature as there is  no ``lake'' of finite 
extent, and the ocean shoreline have no bays.  Another 
property of the Gaussian filtering is that the variance is also monotonically 
decreasing as increasing $ R $.  In general, this picture is also valid in 3-D
random field.  With the help of this feature,
we can distinguish the accretion and the merge in the landscape
produced from initial fluctuation fields.

\subsection{Definition of Monotonic Accretion and Merging in the Landscape} 

In order to understand the monotonic evolution of the field smoothed 
with the Gaussian filter of the
resolution scale $ R $, we rewrite the derivative of the field as  
\bea 
\frac{\partial F(\mbox{\bf r}; R)}{\partial R} = 
R \nabla^2 F(\mbox{\bf r}; R) \; .  
\eea 
This is identical to a diffusion equation of the variables 
($R^2$, $\mbox{\bf r}$).   For all critical points of $ \nabla^2 F <0 $ 
($ \nabla^2 F > 0 $) like peaks (holes), 
it guarantees the monotonically decreasing (increasing) as the scale 
increasing as 
\bea 
\frac{\partial F(\mbox{\bf r}; R)}{\partial R} < 0  \; ,    
(\frac{\partial F(\mbox{\bf r}; R)}{\partial R} > 0)  \; .   
\eea 
This monotonicity of the peak smoothing also 
guarantees that the peak runs continuously on the ridge to 
the shore cape without the island in the landscape.   
It can be reasonable that the smoothing and the merging of peaks
are defined as the continuous accretion growth and the
merging event of bound objects.   
If we have islands in the landscape with other filters 
of Gaussian, however, we cannot distinguish the accretion and the 
merging with the confusion for the scale identification of
the related bound objects.  Fortunately,
we can exclude this problem
as long as using the Gaussian filter which guarantees the absence of
island in the landscape as shown above.  
This is the reason that we take the Gaussian filter in this paper.  

A ridge in the landscape, then, 
represents continuous accretion growth of a bound object.
On the other hand, some ridges terminate on the slope of
neighboring ridges.  
The vanishing point of the ridge on the 
slope of the neighboring ridge can be defined as a type of critical 
points.  We call it as {\it sloping saddle} since
it is a saddle point on the slope of the neighbor peak.
The sloping saddle can represent the 
reasonable feature that a bound 
object loses the identity loss in the merger, associated with
a tree structure in which the branches of the ridges 
are nested at the junctions of the sloping saddles.
Then, we can translate the topology in the landscape to the tree structure
named with {\it skeleton tree}, which consists of
the accretion branches and the merger junction picked up with the
sloping saddles.  

\subsection{Skeleton Tree Picture}

%Fig1 
\begin{figure}
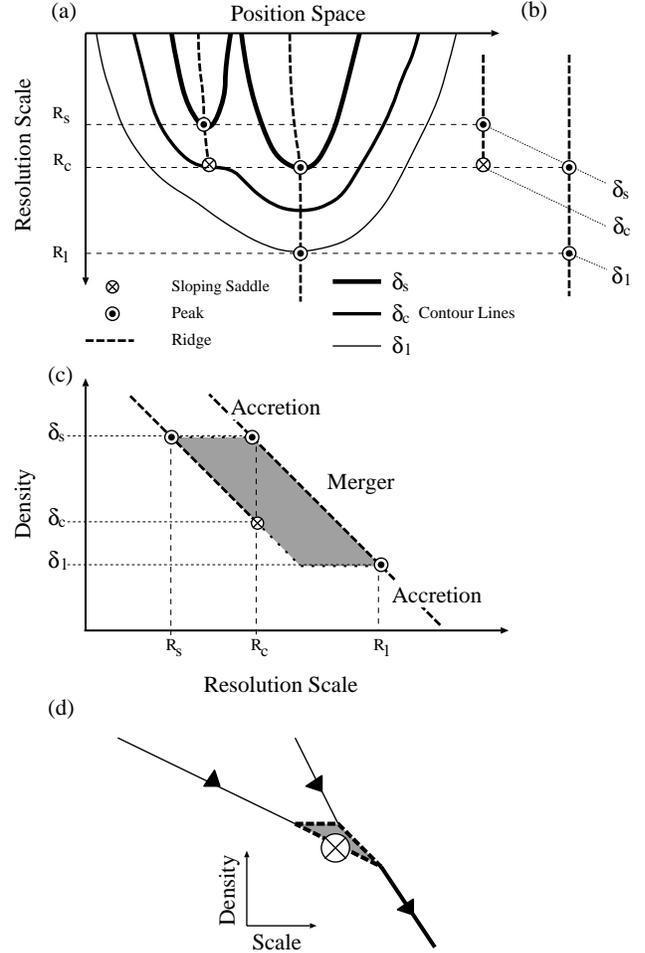

\begin{center}
\psbox[scale=0.45]{fig1.eps}
\end{center}
\caption{ The schematic presentation for the abstraction steps 
from the smoothing fluctuation field to the {\it skeleton tree} picture.
(a)  Schematic representation of the landscape.  
The symbols of $\odot$ and $\otimes$ means the peak and the sloping
saddle, respectively.  The peak trajectory with filter smoothing
makes the ridge.  A ridge terminates at a sloping
saddle at $ (R_c , \delta_c) $ which is associated with a 
ridge of the neighboring peak
through the resolution scale interval from $R_c$
to $R_l$.  (b)  The critical points and the ridges are picked up as
the abstract tree of the field around a sloping saddle.
(c) The tree is presented by the reordering in $ ( R,\delta ) $ space.
The hatched area means the merger.  
(d) The graphical presentation of the smoothing peaks with
the merger and the accretion in {\it the skeleton tree formalism}.  } 
%\lable{mtreef1}
\end{figure}

We will describe the abstraction from the landscape to the tree structure 
as presented in Fig. 1 schematically.
In Fig. 1 (a), the local structure of the landscape is represented 
with contour lines, and the different classes of interest 
critical points are marked;
the peak as $\odot$ and the sloping saddle as $ \otimes $, and   
the ridges are represented as the dashed lines.  
In this example, the sloping saddle appears on a resolution scale 
$ R_c $ at a threshold $ \delta_c $.
A ridge terminates at the sloping saddle
$ \otimes(R_c, \delta_c) $.  It means that a bound 
object of the scale $ R_c $ loses its identity at
the threshold $ \delta_c $.  There is a
ridge neighboring with the sloping saddle, on which the peak has 
the density of $ \delta_s $ at the same resolution scale 
of $ \otimes(R_c, \delta_c) $.  The peak on the neighboring ridge is
reforming around the sloping saddle. Then, 
we can consider that the reformation starts from $ (R_c,
\delta_s) $ of a peak neighboring with the sloping saddle.  
The merger with the destruction and the reformation  
starts from $ \delta_s $ and ends at $ \delta_l $ with the scale
range from $ R_s $ to $ R_l $ in the case of Fig. 1 (a).  

As shown in Fig. 1 (b) as an abstraction of Fig. 1 (a), then, 
we can introduce a tree structure associated with  
the sloping saddle $ \otimes(R_c, \delta_c) $ and three peaks;  
the neighboring peak $ \odot(R_c,\delta_s) $ of $ \otimes(R_c,
\delta_c) $ in the same resolution $ R_c$,  
the progenitor peak $ \odot(R_s,\delta_s) $ on the same ridge of $ \otimes(R_c,
\delta_c) $ at $ \delta_s $ of the density of
$ \odot(R_c,\delta_s) $, and the reformed peak 
$ \odot(R_l,\delta_l) $ on the same ridge of the neighboring
$ \odot(R_c,\delta_s) $.  

For describing the history of the hierarchical clustering with the merging, 
it is convenient to reorder the tree structure along the 
threshold level instead with the filtering scale.  Because the 
threshold level is identical to the time as monotonically mapped
with $ \delta_c=\delta_{c;0} D(t)^{-1} $.  
Fig. 1 (c) schematically shows this reconstructed feature  
from the Fig. 1 (b).  
The continuous accretion growth 
is represented as the dashed lines along the ridges, while 
the merger is represented as the hatched area in Fig. 1 (c).
In this case, the merger occurs during the interval 
between $ \delta_l $ and $\delta_s $.
Even if we can follow the neighboring peak on the
same ridge over $ \delta_c $ in the landscape, however, we have two peaks
at a threshold during the merger after the reordering with the
threshold, as shown in the hatched area in Fig 1. (c).
It means that the reforming peak on the ridge
neighboring with the sloping saddle also 
loses its identity with the merger.  Then,
the neighboring ridge is divided into two part of a disappearing peak
$ > \delta_c $ and a reforming peak $ < \delta_c $.  

The merger feature is simplified as a joint, which connects three
lines of two disappearing peaks  
and the reforming peak at a point of
the sloping saddle as shown in Fig. 1 (d), where the lines  
are the abstraction of the ridges as the continuous accretion growth.
In the Figure, the line processes of the accretion branches
are connected with the point process of the merger
at the $ (R_c,\delta_c) $,
which makes a {\it skeleton tree}.

In the derivation of the {\it skeleton tree},
we have neglected the detail features of three peaks around $ \otimes(R_c,
\delta_c) $ except counting the number of the associated lines,
and approximated as 
the destruction and the reformation occur at $ (\delta_c,R_c) $ 
instantaneously.  
This is reasonable since no other parameter sets
except $ (\delta_c,R_c) $ can be defined without any extra
parameter for modeling the merger event.  The consistency of this
simple picture can be proved with the result in the section 6.  

We can practically reproduce accretion and merging with 
the script of {\it skeleton tree} as shown in in Fig 1(d);
the arrowed line is related to 
the mass growth with continuous accretion, and the joint of hatching
triangle is a merger event with the destruction and the reformation
with the mark $ \otimes $ of the sloping saddle.  
We note that the branching number at the merging 
points is always two in the progenitor side.
This is natural since the merger rate 
of three or more multi peaks are negligible to that of double peaks.  

\section{The Constraints of Critical Points}

\subsection{Line and Point Processes in the Skeleton Tree}

As shown in the previous section,
{\it the skeleton tree formalism} is described by a set of the line and  
point processes of the smoothing peak along the ridge and the sloping saddle
in the extended space $ ({\bf r}, R) $.  The line process of the
smoothing peaks in $ ({\bf r}, R) $ is equal to the point process in the 
original spatial coordinate $ {\bf r} $.  

Their density fields of these point processes 
are described as sums of $ \delta $ functions :  
\bea 
n_{pk} ({\bf r};R) & = & 
\sum_i \delta^{(3)}(\mbox{\bf r}-\mbox{\bf r}_{pk,i}) \;,
\\
{\bf n}_{ss} ({\bf r}, R) & = & \sum_i
\delta^{(3)}(\mbox{\bf r}-\mbox{\bf r}_{ss,i})
\delta ( R - R_{ss,i})  \; ,
\eea
where the subscripts of $ pk$ and $ss$ means peaks and sloping
saddles, $ R_{ss} $ is the resolution scale of the
sloping saddles.  
Since $ n_{pk} d^{3}{\bf r} $ and $ {\bf n}_{ss} d^{3}{\bf r} d R $
are the numbers in 3-D infinite volume of $ d^3 {\bf r} $ and 
4-D infinite volume of $ d^{3}{\bf r} d R $,  
we call the former and the later
as the spatial density and the instantaneous spatial density,
respectively.   

We can express the point processes entirely in terms of the field and
its derivatives with the spatial coordinate $ {\bf r} $
and the resolution scale $R$.  
In the neighborhood around a critical point, with its constraint of
$ \nabla F( {\bf r})\vert_{cr}  = 0 $
, we can expand the field in a Taylor series:   
\bea  
F(\mbox{\bf r}) & \simeq & 
F(\mbox{\bf r}_{cr})  +\frac{1}{2!}
\nabla \otimes \nabla F \vert_{cr} \Delta {\bf r} \Delta {\bf r}
\nonumber \\
& & \makebox[0.5cm]{} + \frac{1}{3!}
\nabla \otimes \nabla \otimes \nabla
F \vert_{cr} \Delta {\bf r} \Delta {\bf r}
\Delta {\bf r} \; ,
\eea
and its derivatives can be also expanded as 
\bea 
\nabla F (\mbox{\bf r})
& \simeq & \nabla \otimes \nabla F \vert_{cr} \Delta {\bf r} \; , \\
\nabla \otimes \nabla F (\mbox{\bf r})
& \simeq & \nabla \otimes \nabla F (\mbox{\bf r}) \vert_{cr} +
\nabla \otimes \nabla \otimes \nabla F \vert_{cr} \Delta {\bf r} \;  , 
\eea
where 
$ \Delta {\bf r} = {\bf r} - {\bf r}_{cr} \; $
and the suffix $ cr $ means the value at the critical point.  

The critical points can be divided into non-degenerate one and 
degenerate one.  The extrema like a peak and hole    
can be categorized into non-degenerate one.       
Provided the condition of the non-degenerate extrema 
$ {\mbox{det} (\nabla \otimes \nabla F) \vert_{cr}  \neq 0} \; $,
Eq. (11) can be rewritten to 
\bea
{\bf r} - {\bf r}_{cr} = ( \nabla \otimes \nabla F \vert_{cr} )^{-1}  
\nabla F (\mbox{\bf r}) \; . 
\eea
Using the second derivatives of the field,  
the number density of the extrema 
can be represented as  
\bea 
n_{ex}& = & 
  \delta^{(3)}( 
  (\nabla \otimes \nabla F)^{-1} \nabla F  ) \nonumber \\
 & = &
\vert  \mbox{det}
(\nabla \otimes \nabla F) \vert \delta^{(3)}(\nabla F) \; . 
\eea 
In order to describe the point process for the extrema,
thus, it is enough to take the terms to the 
order of the second derivatives.

In the degenerate case of $ \nabla \otimes \nabla F \vert_{cr}  = 0 \; $,
however, we cannot describe the displacement vector
only with the first and second derivatives of the field as
the non-degenerate case of Eq. (13).  
The sloping saddle is a kind of the degenerate 
critical points.  

Under the transformation of 
the principal axis in the spatial coordinate, the part of six
components 
related to the second derivatives $ \nabla \otimes \nabla F $ 
in the covariance matrix becomes diagonal.  It means that 
the second derivatives have three eigenvalues as 
\bea 
(F_{11},F_{22},F_{33})= - \sigma_2 (\lambda_1,\lambda_2,\lambda_3) \; , 
 F_{\alpha \beta}=0 \; (\beta \neq \alpha) \; , 
\eea
where $ \sigma_2 $ is the rms of $ \nabla \otimes \nabla F $
( see the definition in Appendix A ).   
All the eigenvalues are not null in the non-degenerate cases.
On the other hand, the degenerate critical points
has one null eigenvalue at least as $ \lambda_3 =0 $, where 
we assumed $ \lambda_1 \geq \lambda_2 \geq \lambda_3 $ for convenience.
This is the reason for the break of the non-degenerate condition as  
$ \mbox{det} \nabla \otimes \nabla F \vert_{cr}  = 0 \; $ 
in the degenerate case.  In general,
a sloping saddle has a neighboring peak at the degenerate direction.  
In this degenerate direction, we cannot take the inversion of
Eq. (11) as $ x_3 - x_{3,cr} = F_{33}\vert_{cr}^{-1} F_3 $ since $
F_{33} \vert_{cr} = 0
$.

In order to describe the point process of the sloping saddles,  
we use the expansions
of $ F_{33} $ at the degenerate direction and $ \nabla^{(2)} F $
for the rest 2-D non-degenerate components 
in a couple of Eq. (12) and Eq. (11) as 
\bea
F_{33} & \simeq & \sum^3_{\beta=1} F_{33\beta}  
\Delta x_\beta \simeq F_{333} \Delta x_3 \; ,  \\
\nabla^{(2)} F
& \simeq & \nabla^{(2)} \otimes \nabla^{(2)} F \Delta {\bf r}^{(2)}
\; ,
\eea
where the suffice of $ (2) $ means the 2-D non-degenerate space.  
The set of the expansions derives
\bea
{\bf r}^{(2)} - {\bf r}_{cr}^{(2)} & = & 
( \nabla^{(2)} \otimes \nabla^{(2)} F \vert_{cr} )^{-1}  
\nabla^{(2)} F (\mbox{\bf r}) \; ,  \\
x_3 -x_{3,ss} & = & F_{333}^{-1} F_{33} \; .
\eea 

Furthermore, we should remember that the sloping saddle is defined
in the extended space with the resolution scale $ R $.  The gradient
$\nabla F (\mbox{\bf r}) $ can be expanded with the derivative of $R$; 
\bea 
\nabla F (\mbox{\bf r})
\simeq \frac{d \nabla F  }{d R} \vert_{ss} (R-R_{ss})
= \nabla^2 \nabla F \vert_{ss}\: R (R-R_{ss}) \; . 
\eea
In the degenerate direction, this gives 
\bea
 R-R_{ss} = \left( \sum_{\alpha=1}^3 F_{3 \alpha \alpha} R
\right)^{-1} F_3 \; .
\eea 

We obtain the constraint for the instantaneous spatial density
of the sloping saddles as 
\bea
{\bf n}_{ss} & = & \sum_i
\delta^{(3)}(\mbox{\bf r}- \mbox{\bf r}_{ss,i}) \delta ( R - R_{ss,i}) 
\nonumber \\
& = &
\vert  \mbox{det}
( \nabla^{(2)} \otimes \nabla^{(2)} F) 
\vert \delta^{(2)}(\nabla^{(2)} F) \nonumber \\
&  & \makebox[1cm]{} \times 
\vert F_{333} \vert \delta (F_{33})  \; 
{ \vert {\sum_{\alpha=1}^3 F_{3 \alpha \alpha} R} \vert}  
\delta (F_3)  \; .
\eea

\subsection{Constraints of Peaks and Sloping Saddles} 

Under the transformation to the diagonal principal axis,  
the peaks require the condition of 
$ \lambda_1 \geq \lambda_2 \geq \lambda_3 \geq 0 $  
which should be added to the above condition of the extrema: Eq. (14).  
With the additional condition of the peaks,  
the constraint for the spatial distributions of the peaks is 
\bea 
C(\mbox{\bf F}^{(10)} \vert \mbox{\rm peaks} ) 
& = &\sigma_2^3 
\vert \lambda_1 \lambda_2 \lambda_3 \vert \prod_{i=1}^3  
\delta(F_{\alpha}) \prod_{\alpha=1}^3 \theta(\lambda_{i}) \; , 
\eea
where $ \theta(\lambda_{i}) $ is the Heaviside function.   

The sloping saddles requires the additional condition of 
$ \lambda_{i}>0 \; (i=1,2) $ as excluding the merger of the
hole.  With the constraint of Eq. (22),
the constraint for the instantaneous spatial distribution of
the sloping saddle can be described as 
\bea 
& & C( \mbox{\bf F}^{(20)} \vert \; \mbox{\rm s.saddles} ) = \sigma_2^2 
\sigma_3^2 R  
 \vert \lambda_1 \lambda_2 w_3 (w_1+w_2+w_3) \vert \nonumber \\
& & \makebox[1cm]{} \times 
\delta(\lambda_3) \prod_{\alpha=1}^3 
\delta(F_{\alpha}) \prod_{i=1}^2 \theta(\lambda_{i}) \; , 
\eea  
where we use $ \sigma_3 w_{\alpha} = F_{3 \alpha \alpha} $.

\section{Probability Distribution Function} 
   
The joint probability distribution 
of n-dimensional random variables with multivariate Gaussian can 
be described as 
\bea 
P( \mbox{\bf  F}^{(n)} ) d \mbox{\bf  F}^{(n)} = \frac{
\exp [ \frac{1}{2} \mbox{\bf  F}^{(n)\:T} \mbox{\bf C}^{(n \times n) -1} 
\mbox{\bf  F}^{(n)} ] }
{ (2\pi)^n \vert det \mbox{\bf C}^{(n)} \vert^{1/2} } 
d \mbox{\bf  F}^{(n)} \; . 
\eea 
The convariance matrix $ \mbox{\bf C}^{(n \times n)} $ 
is defined as the expectation 
value of the direct product of the vector $ \mbox{\bf F}^{(n)} $ : 
\bea 
C^{(n \times n)}_{\alpha \beta}= < F^{(n)}_{\alpha} F^{(n)}_{\beta} > \; ,  
\eea 
where the suffix $ (n)$ of the $ \mbox{\bf F}^{(n)} $ 
is the dimension of the 
parameter space.

In general, 
the conditional probability of the event A
with the constraint event B is given by the Bayes formula
$ P( \mbox{A}\vert
\mbox{B}) = P( \mbox{A}, \mbox{B}) / P ( \mbox{B}) $. 
The joint probability, then, can be expanded with the conditional
probabilities. 
If the parameters in the vector are all Gaussian distributed, we can
directly obtain the covariance matrix of the conditional probability.
There is a general theorem which is extremely useful when
we calculate the above joint probability from the
conditional probabilities.    
For the count of the peaks, we need only the 10 dimension parameters 
$ \mbox{\bf F}^{(10)} = (F, \nabla F, \nabla \otimes \nabla F ) $.
For the count of the sloping saddles, 
we should extend the parameter space to the 20 dimension 
$ \mbox{\bf F}^{(20)} = (F, \nabla F, \nabla \otimes \nabla F, 
\nabla \otimes \nabla \otimes \nabla F) $.
The 20 dimension covariance matrix can be found with the explicit forms of
the parameters in Appendix A.  With the help of the theorem 
(see Appendix B1),  the divided conditional probabilities
are described for the case of the peaks and the sloping saddles
in Appendix B2 and B3, respectively.  

For a type of critical points
in a n-dimensional parameter space described with constraints, in general, 
the probability weighted density is
\bea
n_{cr}({\mbox{\bf F}}^{(n)}) d \mbox{\bf  F}^{(n)}
= P({\mbox{\bf F}}^{(n)}) C({\mbox{\bf F}}^{(n)} \mid 
\mbox{cr. points} ) d \mbox{\bf  F}^{(n)} \; .
\eea 
Then, 
the ensemble averaged density of the type of critical points 
restricted with m-dimension parameters, can be obtained from 
the integration over (n-m) dimension parameters; 
\bea 
< n_{cr}({\mbox{\bf F}}^{(m)}) > d \mbox{\bf  F}^{(m)} 
 = \int d {\mbox{\bf F}}^{(n-m)} 
n_{cr}({\mbox{\bf F}}^{(n)})
d \mbox{\bf  F}^{(m)}
\; . 
\eea
The ensemble averaged density gives the scale function we interest.  
   
\section{Scale Functions of Critical Points} 

We will derive the differential number densities of all the peaks,
the nesting peaks, the non-nesting peaks,  and   
the instantaneous differential number density of the sloping
saddles,    
which are presented as the scale functions and
the instantaneous scale function of 
$ N_{pk}(R,\delta) dR$, $ N_{nest}(R,\delta) dR$,
$ N_{pk}(R,\delta) dR$, and $ {\bf N}_{ss}(R,\delta) dR d \delta $,
respectively.

The scale functions for all the peaks, the non-nesting,
and nesting peaks are briefly
given in the appendix C1, C2, and C3,  according to BBKS and 
the CUSP formalism.  
We had checked that the contribution of the nesting peaks 
is negligible as shown in Fig. 2 .  Thus,
we treat the scale function of peaks as that of the non-nesting peaks
hereafter. 

In the first subsection, then, 
we sketch the derivations of the scale functions of peaks.   
In the next subsection, 
we present the new calculation for the instantaneous scale function
for the sloping saddles.  

\subsection{Scale Function of Peaks}  

%Fig2 
\begin{figure*}
%\begin{center}
\psbox[scale=0.9]{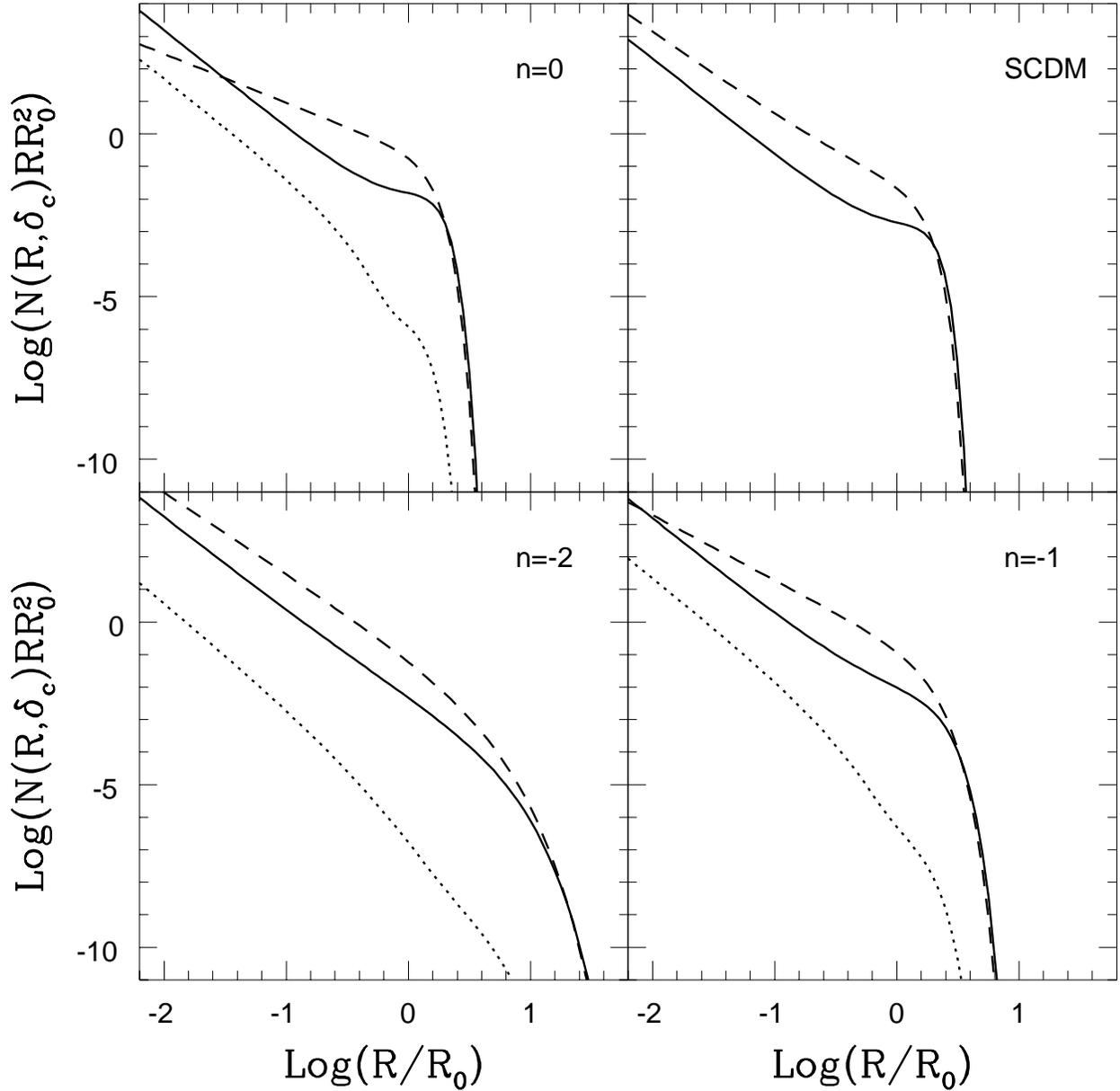}
%\vspace{302pt}
%\end{center}
\caption{ The scale functions are presented
for various fluctuation models of the initial density fluctuation.
The solid and dotted lines represent the scale function of the
all peaks and the nesting peaks.  The dashed line is that of the
PS formalism with the same filter.
}
\end{figure*}

The scale function of $ N_{pk}(R,\delta) dR$  
can be calculated from the ensemble-averaged density of
the peak $ {\cal N}_{pk} 
(\nu; R) d \nu = < n_{pk} (\nu ;R) > d \nu $,
where $ \nu = \delta/ \sigma_0$,
according to Manrique $\&$ Salvador-Sole (1995). 
The scale function means the differential number density
per infinitesimal range of $ R $ at a fixed $ \nu $, not from that 
per infinitesimal range of $ \nu $ at a fixed $ R$.   
The forward one and the last one are denoted 
by a roman capital and a caligraphic capital which is the same
as the notation of BBKS, respectively.   
The spatial density for the peaks of the non-degenerate 
critical points with 
a certain filtering scale can be 
calculated by the same way of BBKS.  

In order to obtain  
the density of peaks with up-crossing a certain $ \delta_c $ 
in the range of the filtering scale between $ R - \Delta R$ and $ R $, 
we should pick up the critical points which are equal 
to or smaller than $ \delta_c $ 
at the filtering scale of $ R $ and becomes larger than $ \delta_c $ at the 
smaller filtering scale of $ R - \Delta R $.  It means that 
we should count the peaks of 
$ \delta_c + (d F_{pk}/ dR) \Delta R < F_{pk} < \delta_c $ 
on scale $ R $ .  As shown in the discussion of the Gaussian filtering, 
the condition of the counting is expressed as 
\bea 
\delta_c + \nabla^2 F  R \Delta R < F < \delta_c  \;  .  
\eea 

As obtained for the peaks in BBKS, in general,   
we can obtain the spatial number density of the peaks 
in infinite ranges $ d \nu dx $ of the density contrast $ \nu $ 
and the second derivative of it with a certain filtering scale $ R $;  
$ {\cal N}_{pk} (\nu,x ;R) d \nu dx $.   We can 
transform this number density to the number satisfying the condition of 
the our counting;  
\bea 
N_{pk}(R,\delta_c) & = & \lim_{\Delta R \rightarrow 0}
 \frac{1}{\Delta R} \int^{\infty}_{0} dx \int^{\nu_{c}}_{\nu_{c,b}} 
d \nu {\cal N}_{pk}(\nu,x ;R) \nonumber \\ 
& = &  {\cal N}_{pk}(\nu_c ;R) <x>_{pk}
\left( \frac{\sigma_2(R)}{\sigma_0(R)} \right) 
R \;, \\ 
\nu_{c,b} & = &\nu_c 
-x \left( \frac{\sigma_2(R)}{\sigma_0(R)} \right) R \Delta R  \; ,
\eea   
where $ < x >_{pk} $ is the mean value of $ x $ defined 
from the distribution function as 
\bea 
< x >_{pk} = \frac{ H_{pk}(\gamma, \gamma \nu_c ) }
{ G_{pk}(\gamma, \gamma \nu_c) } \; , 
\eea
In practical treatment, we can use the averaged mapping relation
\bea
 d \nu_c \simeq
<x>_{pk} \left( \frac{\sigma_2(R)}{\sigma_0(R)} \right) 
R d R \; .
\eea 

Fig. 2 illustrates $ N_{pk}(R,\delta_c) R R_0^2$ and $ N^{nest} (R,
\delta_c) R R_0^2$ versus $ R/R_0 $  with the present
collapsing threshold condition $ \delta_c =1.68 $ in the power-law
fluctuations of $ n= 0, -1 $, and $ -2 $.  The former and the latter
are represented with the solid  and dotted lines, respectively.    
The critical scale of $
R_{0} $, was determined by $ \delta_c/ \sigma_0 (R_0) = 1 $
with $ \delta_c =1.68 $.  
The number of the nesting peaks is
negligible compared with that of all peaks. Hereafter, then, we will 
take $ N (R,\delta_c) = N_{pk}(R,\delta_c) $.  
The PS scale functions, obtained from the same filtered random fields, 
are also presented with the dashed lines.
We also present the two scale functions of peaks for
standard CDM (SCDM) model with the normalization 
of 8 Mpc with the bias factor $ b= 1$.  
Even if 
the two functions in the small scale are different from each other between  
the peak theory and the PS formalism as pointed out by Appel $\&$
Jones (1991),  they deviate relatively little over the cosmological interest
interval ( 2 decades in the filter scale, 6 decades in the mass scale
around $ R_0 $).

Fig. 2 shows that
the scale functions of the peaks $ N_{pk}(R,\delta_c) R R_0^2$ 
become proportional to $ R^{-3}$ asymptotically in the small scale.
As shown in BBKS, the cumulative number is a useful quantity which can 
be evaluated analytically: 
\bea
n_{pk}( \nu_c=- \infty ) = \int^{\infty}_{\nu_c=-\infty} {\cal N}_{pk}
d \nu =
\frac{29-6\sqrt{6}}{ 5^{3/2} 2 (2 \pi)^2 }
R_{\ast}^{-3} \; ,
\eea 
As $ R \rightarrow 0 $, the density contrast $ \nu_c \rightarrow
0 $ and the peaks of the low contrast $ \nu_c $ dominates
in the cumulative number.  Then,
$
N_{pk}(R,\delta_c) R R_0^2 \propto
n_{pk}(\nu_c \rightarrow 0; R_{\ast} ) \propto
n_{pk}(\nu_c=- \infty ; R_{\ast} ) $ in the small scale.   
Since $ R_{\ast}$ is proportional to
the resolution scale $ R $,  the asymptotic feature seen in Fig. 2 
is not unexpected thing.   

\subsection{Instantaneous Scale Function of Sloping Saddles}

The instantaneous scale function of sloping saddles can be
directly calculated as its ensemble-averaged density:
\bea 
\mbox{\bf N}_{ss}(R,\nu_c) d R d \nu_c & = & < {\bf n}_{ss} (\nu_c;R)
> d R d \nu_c \; .
\eea
The brief calculation is described in the appendix C.  

We should note that the scale function is defined as the 
density in the infinite interval $ d R d \nu_c$.  
The scale function of the peaks 
$ N_{pk}(R,\delta_c) dR $ introduced before is the differential density  
of peaks per infinitesimal ranges of the resolution scale $ R $ at a 
fixed density threshold $ \delta_c $. On the other hand,
the instantaneous scale function of the sloping saddles
$ \mbox{\bf N}_{ss}(R,\delta_c) $
is the differential density per infinitesimal ranges 
of $ R $ and $ \delta $.  Their difference come from the fact that
the point processes of peaks and sloping saddles can be defined
in the 3-D spatial coordinate and the 4-D space extended with $ R $,
respectively.  Then, the instantaneous scale function is responsible
for the time evolution properties of the scale function
with the merging process as seen in the {\it skeleton tree};
the line of the smoothing peak connected with other lines at the
junction of the sloping saddle.  In the next section,
using {\it the skeleton tree formalism},  
we will present the evolution of the scale functions with
the distinction between the merger and the accretion according to
{\it the skeleton tree formalism}.   

\section{Evolution with Merger and Accretion}

\subsection{Instantaneous Scale Functions for the Disappearing and
Reforming Peaks}

As discussed in the sketch of the {\it skeleton tree} picture with Fig. 1,
the reformation of a peak can be only identified with the destruction
of two peaks around a sloping saddle.  
In the above section, 
we can obtain the instantaneous number density of
the merger of a scale $ R $ at $ \delta_c $ in the 
infinite interval of $ d \delta_c $.
Thus, with changing between the variables $ \delta_c $ and $ \nu_c $,
the instantaneous scale functions of the destruction and the
reformation for the peaks of the scale $ R $ at $ \delta_c $
are described as 
\bea 
\mbox{\bf N}^d (R,\delta_c) d R \; d \delta_c  & \simeq & 
2 \mbox{\bf N}_{ss}(R,\delta_c) d R \; d \nu_c \; , \\
\mbox{\bf N}^f (R,\delta_c) d R \; d \delta_c  & \simeq & 
\mbox{\bf N}_{ss}(R,\delta_c) d R \; d \nu_c \; , 
\eea 
where the superscripts of $f$ and $d$ mean the reformation and the
destruction in the merger, respectively.  

With the formula,
the net change of the differential number density of the peaks in $ d R $ 
during the interval of $ d \delta_c $ is simplified to  
\bea
S (R,\delta_c) d R \; d \delta_c
& = &  
\mbox{\bf N}^f (R,\delta_c) d R \; d \delta_c - 
\mbox{\bf N}^d (R,\delta_c) d R \; d \delta_c  \nonumber \\
& = & 
- \mbox{\bf N}_{ss}(R,\delta_c) d R \; d \nu_c \; .
\eea
It means the net destruction of the peaks is the same as the
appearance of the sloping saddles.
This is reasonable since the formation of the peaks
does not occur in exact sense and all peaks only disappear at the
sloping saddle in the original filtering process. 
From the concept of the {\it skeleton tree},
the peak neighboring with a sloping saddle
is interpreted to lose its identity with the
merger.  We considered that 
the ridge of the neighboring peak 
is divided into two part of a disappearing peak
$ > \delta_c $ and a reforming peak $ < \delta_c $,
even if the ridge of the neighboring peak continues over $ \delta_c $. Thus,
the reforming number of the peaks can compensate
for the destruction of them, and the appearing number of the sloping saddles
lefts as the net disappearing number of the peaks.
The consistency of this formula can be checked with the conservation
equation for the scale function of the peaks, as shown
in the next subsection.   

\subsection{Instantaneous Rates for Accretion, Destruction and Reformation}  

%Fig3 
\begin{figure*}
%\vspace{302pt}
\begin{center}
\psbox[scale=0.9]{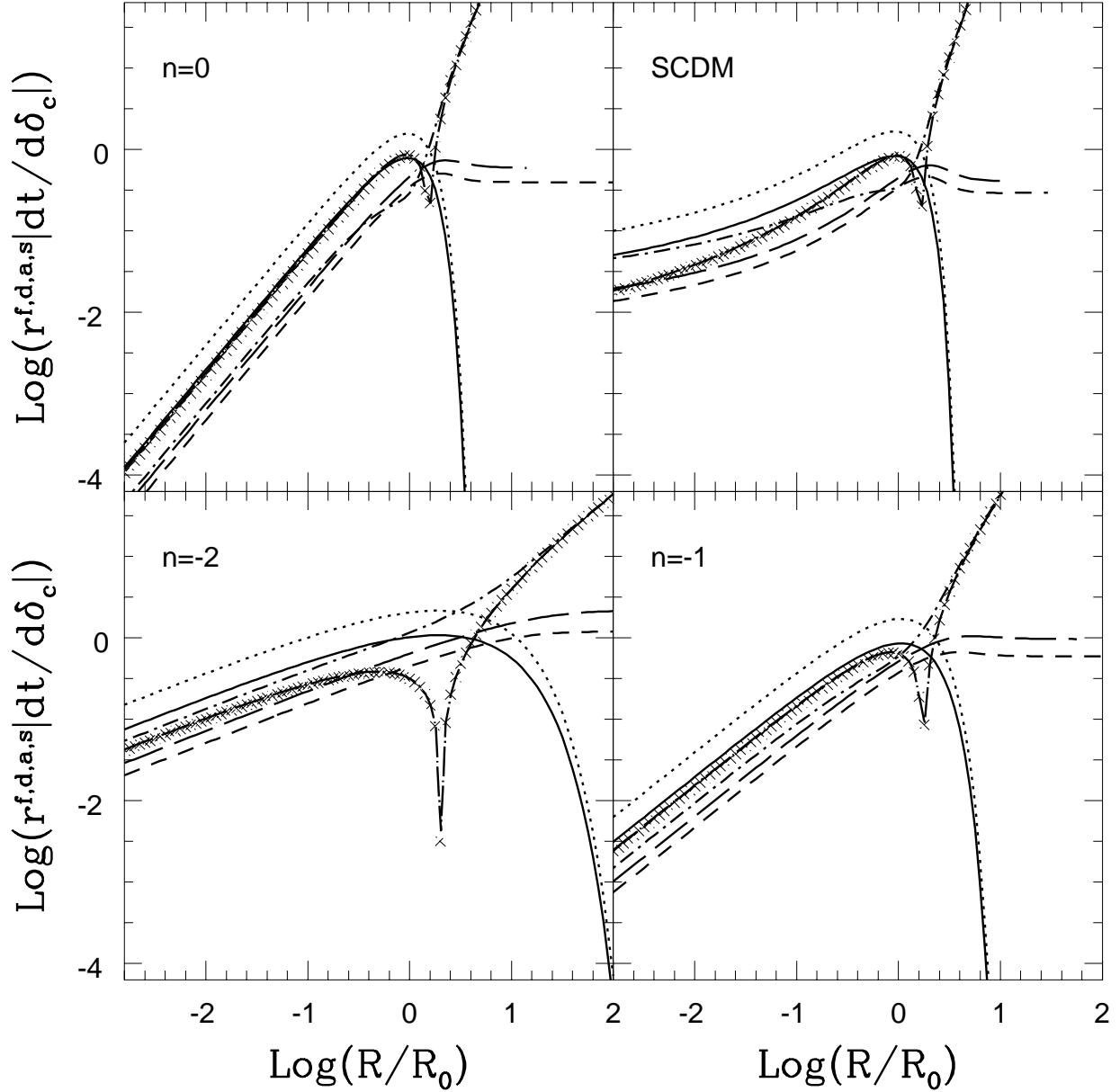}
\end{center}
\caption{ The rates for the evolution processes at the present with
$ \delta_c = 1.68 $. 
The solid, dotted, short dashed, long dashed,
and long dash-dotted 
lines mean the reformation rate $ r^f \vert dt /d \delta_c \vert $,
the destruction rate $ r^d \vert dt /d \delta_c \vert $,
the relative accretion rate $ r^a \vert dt /d \delta_c \vert $,
the relative accretion rate $ r^a \vert dt /d \delta_c \vert $
during the merger, and  
the shift rate $ r^s \vert dt /d \delta_c \vert $, respectively.  
The evolution rate of the scale function 
$ \vert  \partial \ln N_{pk} (R,t)/ \partial t
dt /d \delta_c \vert $ and
the right hand side of the conservation equation (51) is presented
by short dash-dotted line and crosses. The absolute
values are presented for the two last rates.  
}
\end{figure*}

We will consider the instantaneous scale growth rate for a peak of 
$ R $ as the continuous scale growth of the accretion.
The accretion growth rate can be described 
with the scaled Laplacian $ x $ as 
\bea 
{\dot R}_{acc}(R, x, \delta_c ) d t 
& \equiv & \frac{1}{x \sigma_2 R} d \delta_c \; .
\eea
We should note that the growth rate 
for a peak depends on its particular value $ x $.
According to Manrique $\&$ Salvador-Sole (1996),
the mean growth rate for the objects of the scale $ R $ can be expressed as 
\bea 
{\dot R}_{acc}(R, t) d t 
& \equiv &
< \frac{1}{x \sigma_2 R} > d \delta_c \simeq \frac{1}{<x>_{pk}\sigma_2 R}
d \delta_c \; ,  
\eea 
where $ < \; > $
means the average of the function with $ x $ and we used the relation 
\bea 
< \; x^{-1} \;  > & \simeq &
\frac{ \int_0^{\infty} dx \; x^{-1} \; N_{pk} (R, \delta_c, x) dR } 
{ \int_0^{\infty} dx N_{pk} (R, \delta_c, x) dR } \nonumber \\
& = & \frac{ \int_0^{\infty} dx {\cal N}_{pk}(\nu_c, x; R)
\frac{\sigma_2}{\sigma_0} d R } 
{ \int_0^{\infty} dx \; x \; {\cal N}_{pk} (\nu_c, x, R)
\frac{\sigma_2}{\sigma_0} d R } \nonumber \\
& = & \frac{G_{pk} (\gamma, \gamma \nu_c)}
{ H_{pk} (\gamma, \gamma \nu_c)} = <x>_{pk}^{-1}\; . 
\eea
From this mean scale growth rate, we can define 
the relative growth rate with the accretion as
\bea
r^a (R,t) dt & \equiv &\frac{ \partial {\dot R}_{acc} (R,t')}{ \partial R}
\; .
\eea

With the similar consideration applied for the merging objects,
we can introduce the mean scale
growth rate and relative growth rate contributed
with continues accretion during the merger phase;   
\bea 
{\dot R}_{acc,m}(R, t) d t 
& \simeq & \frac{1}{<x>_{ss}\sigma_2 R}
d \delta_c \; ,  \\ 
r^a_m (R,t) dt & \equiv & \frac{ \partial {\dot R}_{acc,m} (R,t')}
{ \partial R} \; ,   
\eea
where
$
<x>_{ss} = H_{ss} (\gamma, \kappa, \gamma \nu_c) /
G_{ss} (\gamma, \kappa, \gamma \nu_c) \;$ .

The instantaneous destruction and reformation rates can be defined directly
from their instantaneous scale functions as 
\bea
r^d (R,t) \; dt & = & \frac{ {\bf N}^d (R, \delta) d R } { N (R,t) d 
R} \; d \delta \; \\
r^f (R,t) \; dt & = & \frac{ {\bf N}^f (R, \delta) d R } { N (R,t) d 
R} \; d \delta \; . 
\eea  
Thus, the conservation equation for the scale function $ N (R,t) $ 
is given as 
\bea
\frac{ \partial N (R,t) }{\partial t} +
\frac{ \partial  ( {\dot R}_{acc} N (R,t) ) }{ \partial R}  =  S(R,t) \; ,
\eea
where $ S(R,t) $ can be given as
the net source term with the destruction and reformation
rates of $ r^d (R,t) $ and $ r^f (R,t) $ ;   
\bea 
S(R,t) = [ r^f (R,t) - r^d (R,t) ] N (R,t) \; . 
\eea
The conservation equation can be rewritten to
\bea 
\frac{ d \ln N (R,t) }{d t} = r^f (R,t) - r^d (R,t) - r^a (R,t)\; , 
\eea
which can be rewritten to
\bea 
\frac{ \partial \ln N (R,t) }{\partial t} = r^s(R,t)  + 
r^f (R,t) - r^d (R,t) - r^a (R,t)\; , 
\eea
where
\bea
r^s(R,t) = -{\dot R}_{acc} \frac{\partial \ln N (R,t)}{\partial R} \; ,
\eea
is the rate of shift in the scale space with the
accretion for the number density distribution.

The inverses of these rates of $ r^{a,f,d,s}(R,t) $ give the time
scales of the individual processes.      
In Fig. 3, the rates for the reformation,
the destruction, the relative growth with accretion, the relative
growth with continuous accretion for merging objects
and the shift in the scale with the accretion 
at the present are presented with solid, dotted, short
dashed, long dashed, and long dash-dotted lines.  The left hand side 
and the right hand side of the conservation equation (52) are also
presented with short dash-dotted line and crosses,  the former
$ \partial \ln N(R,t) \partial t $ can be calculated with the partial
time derivative of $  N_{pk}(R,t) $ and the latter
is calculated from the set of the
individual rates as  $r^s(R,t)  +  r^f (R,t) - r^d
(R,t) - r^a (R,t)$.   

We can see 
that the partial time derivative of $ \partial \ln N(R,t) / \partial
t $ can be reproduced well from  
the right hand side
of the conservation equation from the individual calculations of 
$r^s(R,t), r^f (R,t), r^d (R,t),$  and $ r^a (R,t)$ as shown in Fig. 3.
Especially,  around the scale of $ R \simeq R_0 $,
both of them shows the same feature, in which the values
shapely switch from negative to
positive.  Then, we can verify the consistency of our formalism with
the equality between the left hand side and the right hand side
of the conservation equation.  

The rates of the reformation and the destruction have a maximum
around $ R \simeq R_0 $ and rapidly decrease for the larger
scale.  On the other hand, 
the accretion dominates the merger processes of the reformation and
the destruction and the rate becomes almost constant 
for the large objects ( $ R >> R_0 $ ).
It is consistent with the picture that 
the large object are growing in the cosmic time scale.
These mean that the larger
objects of $ R > R_0 $ grows with the accretion in the linear regime
of the gravitational instability,
while the smaller objects are accumulated into
the larger objects with the merging in the non-linear gravitational
growth regime under the critical scale of $ R_0 $, which is related to 
the present threshold $ \delta_c $.  In the theory of
gravitational instability,  the smaller than $ R_0$ can become in
non-linear growth.  
We should remark that the merger process is the most efficient
around the critical scale $R_0$ even if the merger still dominates
the accretion in the small scale.  
These properties of the dominating merger around $ R_0 $
are already suggested from the previous
N-body calculations (e.g. Navaro, Frenk, $\&$ White 1997 ).  

%Fig4 
\begin{figure}
\begin{center}
\psbox[scale=0.45]{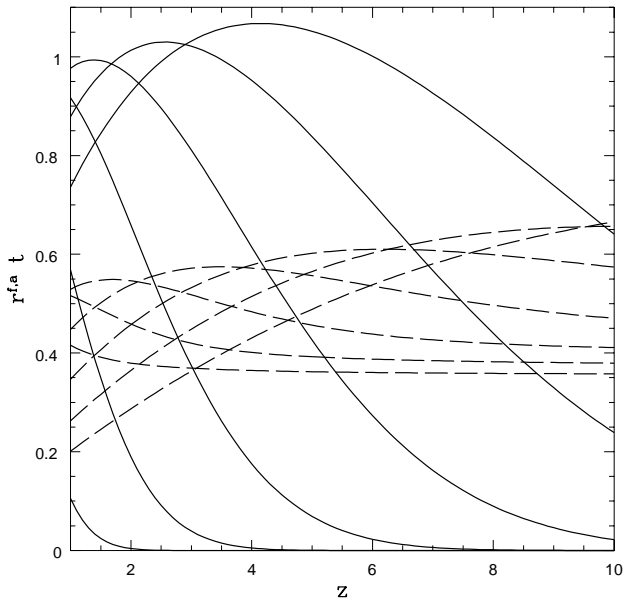}
\end{center}
\caption{ The evolution of growth rates 
at fixed masses for the SCDM model in an Einstein-de Sitter universe.  
The solid and dashed lines present the merger and the accretion
growth rate $ r^f \cdot t $ and
$ r^a \cdot t $.
In the figure, the highest curves at the right is for $ M/M_0 = 
10^{-2} $, and successive curves are for $ M/M_0 = 
10^{-1.5}, 10^{-1}, 10^{-0.5}, 1, 10^{0.5}$ and $ 10^{1}$.   }
%\lable{mtreef1}
\end{figure}

From the view point of the mass growth history for a object of the
scale $ R $,  the results in Fig. 3 should indicate that  
the dominant growth process switches
from the accretion to the merger accumulation around
the threshold of $ \delta
= \sigma_0 (R) $ and
the merger accumulation rate becomes maximum and decreases rapidly.  
We can see directly this feature in Fig. 4, which represents the
evolution of the growth rate with the reformation and the accretion
at fixed scales.
From these results, in general, the growth of a halo 
firstly starts with the accretion process, 
secondly switches to the merging and is suppressed soon after the
merger dominates.  This feature also was remarked from the
simulations.

We should remark that the results in Fig. 4
are not the mean growth histories for the individual objects 
since the fixed scales cannot be directly related to the final 
scales of the objects at the present due to the successive destruction
and the reformation cooperated with the accretion.
In order to reconstruct the mean
growth history for a individual objects, we should extend
the basic of {\it the skeleton tree formalism} including the
background effects.  This extension of {\it the skeleton tree
formalism } will be described in the
following papers.  

\section{Conclusion}

We have derived an analytical expression for the statistics and the
evolution of the cosmic bound objects 
with the reformation, the destruction and the accretion. It is 
applicable to any hierarchical clustering models in which structure
grows via gravitational instability.  {\it The skeleton formalism} 
is derived as a natural expansion of the peaks theory of BBKS.
In the landscape reproduced from the random field with the Gaussian filter, 
we have followed the smoothing of the peaks as
the accretion growth of the objects, and 
have picked up the ``sloping saddles'' as the merging
events with the destruction and the reformation of the objects.
The line and point processes of the peaks and sloping saddles in the
landscape produce the tree structure.  Then, we call our scheme as
{\it the skeleton tree formalism}.   
With {\it the skeleton tree formalism}, we can estimate the rates
of the reformation, destruction, and the accretion
in any hierarchical clustering models.
The set of these rates 
can reproduce the evolution of the scale function of the
objects with the conservation equation.  This reproduction of
the evolving scale function verifies the self-consistency of
{\it the skeleton tree formalism}.
With the rate calculation of the individual processes, we can find 
the merger processes are efficient around the critical scale of $ R_0
$ determined as $ \delta_c = \sigma_0(R_0) $.
The dominant growth process of the
objects switches from the accretion to the merger accumulation  
around the critical threshold related to the scale.
It is important to reproduce the
mass growth history with distinguish between the accretion and the
merger when we try to reproduce the cosmic structure and the galaxy formation
in the hierarchical scenario.

\section*{acknowledgments}
The author is grateful to H.J. Mo and S.D.M. White  
for warm hospitality and the discussions 
when H. H. visited in MPA in Garching.  This works was partially 
supported by a Grant-in-Aid from Japanese Ministry of Education, 
Science and Culture.

\begin{appendix}

\section{The Covariance Matrix} 

We will introduce 20-dimension vector 
$ \mbox{\bf  F}^{(20)} = 
( F,\nabla F,\nabla \otimes \nabla F, \nabla \otimes \nabla \otimes \nabla F,
 ) $.  
The $ \nabla F,
\nabla \otimes \nabla F$, and $\nabla \otimes \nabla
\otimes \nabla F $ have three, six and ten independent components,
respectively.  They can be expressed in terms of Fourier transforms: 
\bea 
F(\mbox{\bf r};R) 
& = & \int d^3 \mbox{\bf k} e^{i {\bf k} \cdot {\bf r} } F(k;R) 
\;, \\
F_{\alpha}(\mbox{\bf r};R) 
& = &\frac{\partial F(\mbox{\bf r};R)}{\partial x_{\alpha}} 
=i \int d^3 \mbox{\bf k} e^{i {\bf k} \cdot {\bf r}} k_{\alpha} F(k;R) 
\;, \\
F_{\alpha \beta}(\mbox{\bf r};R) 
& = & \frac{\partial^2 F(\mbox{\bf r};R)}{\partial x_
\alpha \partial x_\beta} \nonumber \\
& = & -\int d^3 \mbox{\bf k} e^{i {\bf k} \cdot {\bf r}} 
k_{\alpha} k_{\beta} F(k;R) \; , \\
F_{\alpha \beta \gamma}(\mbox{\bf r};R) 
& = & \frac{\partial^3 F(\mbox{\bf r};R)}{\partial x_
\alpha \partial x_\beta \partial x_\gamma}  \nonumber \\
& = & -i \int d^3 \mbox{\bf k} e^{i {\bf k} \cdot {\bf r}} 
k_{\alpha} k_{\beta} k_{\gamma} F(k;R) \; . 
\eea 
It is useful to introduce the integrals over the 
filtered fluctuation spectrum: 
\bea 
\sigma_j^2 (R) = 4 \pi \int_0^{\infty} d k k^{2j+2} P(k ; R) \; , 
\eea 
when transforming the above 
values to the non-dimensional ones.   

The covariance matrix $ \mbox{\bf C}^{(20)} $ is defined as the expectation 
value of the direct product of the vector $ \mbox{\bf F}^{(20)} $ : 
\bea 
C_{ij}^{(20)}= <F^{(20)}_i F^{(20)}_j> \; .  
\eea 
We can represent the matrix as 
\bea 
 \mbox{\bf C}^{(20)} & = & \pmatrix{ 
\sigma_0^2 & \mbox{\bf 0} & \mbox{\bf M}_{02}^T & \mbox{\bf 0}
\cr
\mbox{\bf 0} & \mbox{\bf M}_{11} & \mbox{\bf 0} & \mbox{\bf M}_{13}^T 
\cr
\mbox{\bf M}_{02} & \mbox{\bf 0} & \mbox{\bf M}_{22} & \mbox{\bf 0} 
\cr
\mbox{\bf 0} & \mbox{\bf M}_{13}  & \mbox{\bf 0} & \mbox{\bf M}_{33} 
\cr
}\; , \\
\mbox{\bf M}_{11} & = & \frac{\sigma_1^2}{3} \mbox{\bf I}^{(3 \times
3)} \;  , \\ 
\mbox{\bf M}_{22} & = & \frac{\sigma_2^2}{5} \overline{\mbox{\bf M}}_{22} \; , 
\; \;   \\
\overline{\mbox{\bf M}}_{22} & = & 
\pmatrix{
1   & 1/3 & 1/3 &         \cr 
1/3 & 1   & 1/3 & \mbox{\bf 0} \cr
1/3 & 1/3 & 1   &         \cr
    & \mbox{\bf 0} &   & 1/3\mbox{\bf I}^{(3 \times 3)}  
\cr} \; , \\ 
\mbox{\bf M}_{33} & = & \frac{\sigma_3^2}{7}
\pmatrix{
 \overline{\mbox{\bf M}}_{33} & \mbox{\bf 0} & \mbox{\bf 0} & \mbox{\bf 0} \cr 
\mbox{\bf 0} & \overline{\mbox{\bf M}}_{33} & \mbox{\bf 0} & \mbox{\bf 0} \cr 
\mbox{\bf 0} & \mbox{\bf 0} & \overline{\mbox{\bf M}}_{33} & \mbox{\bf 0} \cr 
\mbox{\bf 0} & \mbox{\bf 0} & \mbox{\bf 0} & 1/15               \cr}
\; , \; \; \\
\overline{\mbox{\bf M}}_{33} & = & \pmatrix{ 
1   & 1/5  & 1/5  \cr 
1/5 & 1/5  & 1/15 \cr
1/5 & 1/15 & 1/5  \cr
 } \; , \\ 
\mbox{\bf M}_{02} & = & -\frac{\sigma_1^2}{3}
(\mbox{\bf K}^{(1 \times 3)},\mbox{\bf 0}^{(1 \times 3)}) \; , \\ 
\mbox{\bf M}_{13} & = & -\frac{\sigma_2^2}{5}
(\overline{\mbox{\bf M}}_{13}^{1},\overline{\mbox{\bf M}}_{13}^{2},
\overline{\mbox{\bf M}}_{13}^{3},
\mbox{\bf 0}^{(3 \times 1)}) \;, \; \; \\ 
\overline{\mbox{\bf M}}_{13}^{l} & = & \mbox{\bf e}_l^T \otimes
(1,1/3,1/3) \; , \; \; 
\mbox{\bf e}_i = ( \delta_{1i}, \delta_{2i},\delta_{3i}) \; , 
\eea
where 
$ \mbox{\bf K}^{(n \times m)},\mbox{\bf I}^{(n \times n)}$ and 
$\mbox{\bf 0}^{(n \times m)} $ 
is the $(n \times m)$ dimension matrix with every entry unity, 
the $(n \times n)$ dimension unit matrix 
and the $(n \times m)$ dimension null matrix, respectivelly.  

\section{Calculations of Joint Probabilities}

\subsection{Theorem for the Joint Probability Expansion}

According to Adler (1981) and BBKS, all Gaussian distributed parameters 
represented by the n-dimension vector  
$ \mbox{\bf Z}=( \mbox{\bf Y}, \mbox{\bf X} ) $, which can be devised into m-dimension vector 
$ \mbox{\bf Y} $ and (n-m) dimension vector $ \mbox{\bf X} $, 
imply us that the conditional 
probability $ P( \mbox{\bf Y} \mid \mbox{\bf X} ) $ is a Gaussian 
\bea 
P(\mbox{\bf Y} \mid \mbox{\bf X}) = \frac{\exp[ -\frac{1}{2} \overline{ 
\Delta \mbox{\bf Y}}^T 
< \overline{\Delta \mbox{\bf Y}} \otimes \overline{\Delta \mbox{\bf Y}} 
\mid \Delta\mbox{\bf X} >^{-1} 
\overline{\Delta \mbox{\bf Y}}]
                         }
{ (2 \pi)^{m/2}
\vert \det <\overline{\Delta \mbox{\bf Y}} \otimes 
\overline{\Delta\mbox{\bf Y}} 
\mid \overline{\Delta\mbox{\bf X}}> \vert^{1/2} } \; . 
\eea 
where $ \overline{\Delta \mbox{\bf Y}} = \Delta\mbox{\bf Y} 
- < \Delta\mbox{\bf Y} 
\mid \Delta \mbox{\bf X} > $,  
with the mean;  
\bea 
< \Delta\mbox{\bf Y} \mid \Delta\mbox{\bf X} > & = & < \Delta\mbox{\bf Y} 
\otimes \Delta\mbox{\bf X} >
< \Delta\mbox{\bf X}^2 >^{-1}
\Delta\mbox{\bf X} \; , \\
< \Delta\mbox{\bf X}^2 >^{-1} & = &
< \Delta\mbox{\bf X} \otimes \Delta\mbox{\bf X} >^{-1} \; ,
\eea 
and the conditional covariance matrix: 
\bea 
& & \mbox{\bf C}(\mbox{\bf Y} \vert \mbox{\bf X}) = 
< \overline{\Delta \mbox{\bf Y}} \otimes \overline{\Delta \mbox{\bf Y}} 
\mid \Delta \mbox{\bf X} > \nonumber \\ 
& & \makebox[1cm]{} = 
<\Delta \mbox{\bf Y} \otimes \Delta \mbox{\bf Y}> \nonumber \\
&   & -< \Delta \mbox{\bf Y} 
\otimes \Delta \mbox{\bf X} >  < \Delta\mbox{\bf X}^2 >^{-1}
< \Delta \mbox{\bf X} \otimes \Delta \mbox{\bf Y} > \; , 
\eea 
where $ < \; > $ represents the mean value and 
$ \Delta \mbox{\bf X} = \mbox{\bf X} -< \mbox{\bf X} > $.  
With the theorem, we can obtain the conditional 
probabilities from the covariance matrix.   

\subsection{Divided Conditional Probability} 

We can 
expand the joint probability distribution with the conditional 
probabilities as   
\bea
& &{P(F,F_\alpha,F_{\alpha \beta},F_{\alpha \beta \gamma}) 
dF dF_{\alpha}^{(3)} dF_{\alpha \beta}^{(6)} 
dF_{\alpha \beta \gamma}^{(10)} } = \nonumber \\ 
& & \makebox[1.5cm]{} \; \; 
P(F_{\alpha \beta \gamma} \vert F,F_\alpha,F_{\alpha \beta})  
dF_{\alpha \beta \gamma}^{(10)} \nonumber \\
& & \makebox[2.cm]{} \; \; \times
P(F,F_\alpha,F_{\alpha \beta })dFdF_{\alpha}^{(3)} dF_{\alpha \beta}^{(6)}  \\ 
& &{ P( F,F_\alpha,F_{\alpha \beta}) dFdF_{\alpha}^{(3)} dF_{\alpha
\beta}^{(6)}  
 = }\nonumber \\  
& & \makebox[1.5cm]{} \; \; 
P(F_{\alpha \beta } \vert F,F_\alpha)dF_{\alpha \beta}^{(6)}
P(F,F_\alpha)dFdF_{\alpha}^{(3)}   \; .
\eea
Using the theorem above, these
conditional probability functions are represented in the explicit
forms:    
\bea 
P(F,F_\alpha) & = & \frac{3^{3/2}\exp[ -\frac{1}{2} Q(F,F_\alpha)]}
                         {(2 \pi)^2 \sigma_1^3 \sigma_0} \; , \\
P(F_{\alpha \beta } \vert F,F_{\alpha}) & = & 
      P (F_{\alpha \beta } \vert F) \nonumber \\
& = &  
\frac{ \exp [ -\frac{1}{2} Q(F_{\alpha \beta} \vert F) ] }
     {(2 \pi)^3 \vert det \mbox{\bf C}(F_{\alpha \beta} \vert F)
\vert^{1/2}} \; ,\\     
P(F_{\alpha \beta \gamma}  \vert F,F_{\alpha},F_{\alpha \beta})  
& = & P(F_{\alpha \beta \gamma} \vert F_{\alpha}) \nonumber \\
& = & 
\frac{\exp [-\frac{1}{2} Q(F_{\alpha \beta \gamma} \vert F_{\alpha}) ]}
     {(2 \pi)^5 
      \vert det \mbox{\bf C}(F_{\alpha \beta \gamma} \vert F_{\alpha}) 
      \vert^{1/2} } \; , 
\eea 
where 
\bea
Q(F,F_\alpha) & = &  \frac{F^2}{\sigma_0^2} 
             + \frac{ 3 F_{\alpha} F_{\alpha} }{\sigma_1^2} \; , \\  
Q(F_{\alpha \beta} \vert F) & = &   
            \tilde{F}_{\alpha\beta}^T
            \mbox{\bf C}(F_{\alpha \beta} \vert F)^{-1}
            \tilde{F}_{\alpha\beta} \; , \\
Q(F_{\alpha \beta \gamma} \vert F_{\alpha}) & = & 
 \tilde{F}_{\alpha\beta\gamma}^T 
 \mbox{\bf C}(F_{\alpha \beta \gamma} \vert F_{\alpha})^{-1}
 \tilde{F}_{\alpha\beta\gamma} \; , \\
\tilde{F}_{\alpha\beta} & = &
F_{\alpha\beta}-\mbox{\bf M}_{02}\sigma_0^{-2} F \; ,\\
\tilde{F}_{\alpha\beta\gamma} & = &
F_{\alpha\beta\gamma}-3 \sigma_1^{-2}\mbox{\bf M}_{13}F_{\alpha} \; ,
\eea
and the conditional covariance matrix;  
\bea
\mbox{\bf C}(F_{\alpha \beta} \vert F)  & = & 
       \mbox{\bf M}_{22} - \mbox{\bf M}_{02}^T\sigma_0^{-2}\mbox{\bf
M}_{02}  \nonumber \\
& = & \frac{\sigma_2^2}{5} ( \overline{\mbox{\bf M}}_{22} -
\overline{\mbox{\bf M}}_{202} )         
\; , \\
\overline{\mbox{\bf M}}_{202} & = & \frac{5\sigma_1^4}{9\sigma_2^2\sigma_0^2}
\pmatrix{ \mbox{\bf K}^{(3 \times 3)} & \mbox{\bf 0}^{(3 \times 3)} \cr
          \mbox{\bf 0}^{(3 \times 3)} & \mbox{\bf 0}^{(3 \times 3)}
\cr   } \; , \\
\mbox{\bf C}(F_{\alpha \beta \gamma} \vert F_{\alpha}) & = &  
       \mbox{\bf M}_{33} - \mbox{\bf M}_{13}^T 3 \sigma_1^{-2} 
\mbox{\bf M}_{13} \; .  
\eea

\subsection{Probability Function for Peaks}

For the peaks, the original covariance matrix 
for the probability is represented 
in the 10 dimension vector $ \mbox{\bf F}^{(10)} $.
As presented by BBKS, 
we can see that the three degrees of freedom 
related to the directional dependence drop off 
from the parameter space for the homogeneous fields.  
After introduce the eigen coordinate, 
we can represent the peak number 
density in 7 dimension parameter space.

For convenience, we will introduce a new set of variables 
\bea 
\sigma_0 \nu & = & F \; , \\ 
\sigma_1 \nu_{\alpha} & = & F_{\alpha} \; , \\ 
\sigma_2 x & = & -\nabla^2 F = \lambda_1 + \lambda_2 + \lambda_3 \; , \\ 
\sigma_2 y & = & \frac{(\lambda_1 -\lambda_3)}{2} \; , \\ 
\sigma_2 z & = & \frac{(\lambda_1 -2\lambda_2 +\lambda_3)}{2} \; , \\  
d F_{\alpha \beta}^{6} & = & 
\vert (\lambda_1-\lambda_2)(\lambda_2-\lambda_3)(\lambda_1-\lambda_3) \vert 
d\lambda_1 d\lambda_2 d\lambda_3 \nonumber \\
& & \makebox[1cm]{} \times d \; \mbox{\rm vol}[ SO(3) ] \nonumber \\ 
& = &  \vert 2y(y^2-z^2) \vert \frac{2}{3} \sigma_2^3 dx dy dz 
\; d \; \mbox{\rm vol}[ SO(3) ] \; , 
\eea  
where $ d \; \mbox{\rm vol}[ SO(3) ] $ is the volume element 
of the three-dimensional 
rotational group $ SO(3) $.   
We had used the relations : 
\bea 
\lambda_1 & = & \frac{\sigma_2}{3}(x+3y+z) \; , \\ 
\lambda_2 & = & \frac{\sigma_2}{3}(x-2z) \; , \\ 
\lambda_3 & = & \frac{\sigma_2}{3}(x-3y+z) \; , \\ 
\prod_{ i=1}^3 
d \lambda_{i} & = & \frac{2}{3} \sigma_2^3 dx dy dz \; . 
\eea  

The probability function in these variables of  
$(F,F_{\alpha}, F_{\alpha \beta})$ is represented as that in the 
variables of $ ( \nu, \eta, x, y, z )$ with
$ \eta = (\nu_1,\nu_2,\nu_3) $. 
\bea 
& & P(F,F_{\alpha},F_{\alpha \beta})   
dF d^3 F_{\alpha} d^6 F_{\alpha \beta} =  \nonumber \\ 
& & \; \; P( \nu, \eta, x, y, z,
\overline{\alpha}, \overline{\beta}, \overline{\gamma} ) 
d\nu  d^3 \eta \nonumber \\
& & \mbox{} \; \; \times
\vert 2y(y^2-z^2) \vert \frac{2}{3} 
\sigma_2^{3}
dx dy dz \; \; d \; \mbox{\rm vol}[ SO(3) ] \; ,
\eea 
where $ \overline{\alpha}, \overline{\beta}, \overline{\gamma} $ are the 
Euler's angles.  
The directional dependence is not so important in the isotropic field.
Then the mean 
of peak number density is enough to consider the 
statistics of the density field.  The mean can be obtained by the 
angle independently integrations.  
\bea
& & \int d \; \mbox{\rm vol}[ SO(3) ] 
P( \nu, \eta, x, y, z, 
\overline{\alpha}, \overline{\beta}, \overline{\gamma} ) 
 \nonumber \\
& & \; \; = \frac{2\pi^2}{3!}P( \nu, \eta, x, y, z) \nonumber \\
& & \; \; = \frac{2\pi^2}{3!}P(x,y,z \vert \nu)P(\nu,\eta) \; , 
\eea
where we used $\int d \; \mbox{\rm vol}[ SO(3) ] = 2 \pi^2 /3! $ 
with the 
degenerate of the triad orientation of the axis for the eigenvalues.  

The determinant and the inverse of the conditional covariance matrix 
are given explicitly as  
\bea 
\vert det \mbox{\bf C}( \nu_{\alpha \beta} \vert \nu) \vert^{1/2} = 
      \frac{2}{5^{5/2} \cdot 3^{3}} (1-\gamma^2)^{1/2} \; , \;  
      \gamma = \frac{\sigma_1^2}{\sigma_2 \sigma_0} \; ,  
\eea
where 
\bea 
\mbox{\bf C}( \nu_{\alpha \beta} \vert \nu )^{-1} 
  & = & 
    \pmatrix{ \frac{1}{1-\gamma^2} \mbox{\bf K}^{(3 \times 3)}+\mbox{\bf R} 
  & 
    \mbox{\bf 0} \cr              
    \mbox{\bf 0} & -15 \mbox{\bf I}^{(3 \times 3 )} }  \; ,  \\
\mbox{\bf R} & = & 5 \pmatrix{ 1   & -1/2 & -1/2 \cr 
                  -1/2 &  1   & -1/2 \cr 
                  -1/2 & -1/2 &  1      } \; . 
\eea 

The probability functions $ P(x,y,z \vert \nu), P(\nu)$ 
are the same of that described by BBKS as 
\bea 
P(\nu,\eta) & = & \frac{ 3^{3/2} }{(2\pi)^2} 
\exp[ -\frac{1}{2}( \nu^2 + 
3 \eta \cdot \eta ) ] \; , \\
P(x,y,z \vert \nu) & = & \frac{3^{3} \cdot 5^{5/2}}
{(2 \pi)^3 \cdot 2 \cdot (1-\gamma^2)^{1/2} }  \nonumber \\
& & \mbox{} \; \; \times \exp[-\frac{1}{2}Q(x,y,z)] \; ,  \\ 
Q(x,y,z) & = & \frac{(x-x_\ast)^2}{(1-\gamma^2)} 
+ 15 y^2 +5 z^2 \; ,  x_\ast = \gamma \nu \; .  
\eea 

\subsection{Probability Function for Sloping Saddles}

For the sloping saddles, in general, 
we should start from the original covariance matrix in the 20 dimension
vector with 10 variables of the third 
derivatives of the field added to that for the peaks.   
After introduce the eigen coordinate, however, 
we need the third derivatives in only degenerate direction 
as the additional variables; $ \nu_{3 \alpha \alpha} =  
F_{3 \alpha \alpha}/\sigma_3 $.   
As similar as shown in the case of the peak, finally, 
the statistics of the sloping saddles of the homogeneous field can be 
determined in the parameter space 
$ (\nu,\eta,x,y,z,{\bf w}) $, where $ {\bf w} =
(w_1, w_2, w_3) = (\nu_{311},\nu_{322},\nu_{333}) $.  
The conditional probability with the third derivatives of  
$ {\bf w} $
can be represented as  
\bea
P( {\bf w} \vert \nu_{3}) & = & 
\frac{\exp [-\frac{1}{2} Q( {\bf w} \vert \nu_{3}) ]}
     {(2 \pi)^{3/2} 
      \vert det \mbox{\bf C}( {\bf w} \vert \nu_{3})
      \vert^{1/2} } \; , \\
Q( {\bf w} \vert \nu_{3}) & = & 
\overline{{\bf w}}^T 
 \mbox{\bf C}(\overline{{\bf w}} \vert \nu_{3}) 
\overline{{\bf w}}
\; , \\ 
\overline{{\bf w}} & = &
{\bf w} -
\frac{-3}{5} \kappa \overline{\mbox{\bf M}}_{13} \nu_{3} \; , \\ 
\kappa & = & \frac{\sigma_2^2}{\sigma_3 \sigma_1} \; . 
\eea 
The determinant and the inverse of the conditional covariance 
matrix for the variables $ \overline{\nu}_{3 \alpha \alpha}$  
are represented as 
\bea
\vert det \mbox{\bf C}( {\bf w} \vert \nu_{3}) 
\vert^{1/2} 
 & = & \frac{2}{5^{3/2} \cdot 3 \cdot 7}(1-\kappa^2)^{1/2}
\; , \\ 
\mbox{\bf C}( {\bf w} \vert \nu_{3})^{-1} 
 & = & \frac{1}{(1-\kappa^2)} 
\pmatrix{ c_{11}  & c_{o} & c_{o} \cr  
          c_{o} & c_{22}  & c_{o} \cr  
          c_{o} & c_{o} & c_{33} }
\; , \\
c_{11} & =& 10-7\kappa^2 \;, \\
c_{22} & =& c_{33} = 45 -42\kappa^2 \;, \\  
c_{o} & =& \frac{21\kappa^2 -15}{2} \;.
\eea

With the condition of the first derivatives $ F_3 $ for 
the sloping saddles, we can represent the exponent part of 
the conditional probability:
\bea
& & Q( {\bf w} \vert \nu_3 )  =  
Q_1( w_1 \vert w_2, w_3, \nu_{3} ) \nonumber \\
& & \makebox[1.5cm]{} + 
Q_2( w_2 \vert w_3, \nu_{3} ) +   
Q_3( w_3 \vert \nu_{3} ) \; ,
\eea
\bea
& & Q_1( w_1 \vert w_2, w_3, \nu_{3} )
=  \frac{3(15-14\kappa^2)}{(1-\kappa^2)} 
\nonumber \\
& & \makebox[1.5cm]{} \;\; \times 
\left[ \overline{w}_{1} 
- \frac{(5-7\kappa^2)}{2(15-14 \kappa^2)} 
 \overline{w}_{3}+\overline{w}_{2} \right]^2 \; , \\
& & Q_2( w_2 \vert w_3, \nu_{3} )  =
\frac{3 \cdot 5 \cdot 7 (25-21\kappa^2)}{2^2 (15-14\kappa^2)} 
\nonumber  \\
& & \makebox[1.5cm]{} \;\; \times  
\left[ \overline{w}_{2} 
- \frac{(5-7\kappa^2)}{(25-21 \kappa^2)} 
 \overline{w}_{3} \right]^2 \; , \\ 
& & Q_3( w_3 \vert \nu_{3} ) =   
 \frac{5^2 \cdot 7}{(25-21\kappa^2)}\overline{\nu}_{333}^2 \; . 
\eea

We will list up some integrations as 
\bea
I_1 & = & \int^{\infty}_{-\infty} d w_{1} \exp \left[ -
\frac{1}{2} Q_1( w_1 \vert w_2, w_3, \nu_{3} ) \right] \nonumber \\
& = &
\frac{ \sqrt{ 2 \pi } (1-\kappa^2)^{1/2} }
{ 3^{1/2} (15-14\kappa^2)^{1/2} }  \; ,  \\
I_2 & = & \int^{\infty}_{-\infty} d w_{2} \exp \left[ -
\frac{1}{2} Q_2( w_2 \vert w_3, \nu_{3} ) \right] \nonumber \\
& = &
\frac{ 2 \sqrt{ 2 \pi } (15-14\kappa^2)^{1/2} }
{ ( 3 \cdot 5 \cdot 7)^{1/2} (25-21\kappa^2)^{1/2} }  \; ,  \\
I_{22} & = &
\int^{\infty}_{-\infty} d w_{2} \; w_{2} \exp \left[ -
\frac{1}{2} Q_2( w_2 \vert w_3, \nu_{3} ) \right] \nonumber   \\
& = & \int^{\infty}_{-\infty} d w_{2}
\frac{(5-7\kappa^2)}{(25-21 \kappa^2)} \overline{w}_{3}
\exp \left[ - \frac{1}{2}
Q_2( w_2 \vert w_3, \nu_{3} ) \right] \nonumber   \\
& = & \frac{(5-7\kappa^2)}{(25-21 \kappa^2)} \overline{w}_{3}
I_2 \; .
\eea 
As shown in the definition of the sloping saddles, 
we are interested in mainly the probability for 
$ w_{3} $ in the third derivative components.  
The conditional probability function of $ w_{3} $ 
can be obtained from the integration with the other
third derivative components 
$ w_{1}, w_{2} $  as 
\bea 
P( w_{3} \vert \nu_{3} ) d w_{3} & = & 
\int^{\infty}_{-\infty} d w_{1} \int^{\infty}_{-\infty} d w_{2} 
P( {\bf w} \vert \nu_{3}) \; d w_{3}
\nonumber \\  
& = & \frac{5^{3/2} \cdot 3 \cdot 7}{2 (2 \pi)^{3/2}(1-\kappa^2)^{1/2} } 
\times I_1 \times I_2 \; \nonumber \\
& &  \makebox[1cm]{} \times
\exp \left[ -\frac{1}{2} \overline{Q}( w_{3}) \right] d w_{3} 
 \nonumber \\  
& = & \frac{ 5 \cdot 7^{1/2}} { \sqrt{2 \pi (25-21\kappa^2)}}
\nonumber \\
& & \; \times
\exp \left[ -\frac{1}{2} \overline{Q}( w_{3}) \right] d w_{3} \;, \\ 
\overline{Q}( w_{3} ) & = & \frac{ 5^2 \cdot 7}{(25-21\kappa^2)} 
( w_{3} + \frac{3}{5}\kappa \nu_3)  
^2  \; . 
\eea
We can see the 
reasonable normalization in the integration of
$ P ( w_{3} \vert \nu_3 ) $ with $ w_{3} $ as 
\beq
\int^{\infty}_{-\infty} d w_{3} P ( w_{3} \vert \nu_3 ) = 1  \; . 
\eeq
For the convenience, we will represent the probability of $ w_3 $
with $ \nu_3 = 0 $; 
\bea 
P ( w_{3} \vert \nu_3=0 )
& = & \frac{1}{\sqrt{2 \pi} \sigma_w} 
\exp \left[ - \frac{1}{2}\frac{w_3^2}{\sigma_w^2}\right] \; ,  \\ 
\sigma_w^2 & = & \frac{(25-21\kappa^2)}{ 5^2 \cdot 7} \; .  
\eea 

We will calculate the integration required in the
calculation of the instantaneous scale function of sloping saddles;
\bea 
& & \int d^3 {\bf w} \vert w_3 (w_1 +w_2 +w_3)\vert  
 P ({\bf w} \vert \nu_3 ) = \nonumber \\
& & \makebox[2cm]{}  < w_3^2>_{ss}(1 + q_{ss} )  \; ,
\eea
\bea 
<w_3^2>_{ss} & = & \int^{\infty}_{-\infty} d w_3 (w_3)^2  P(w_3 \vert \nu_3 )
\nonumber \\  
             & = & \frac{(25-21 \kappa^2)}{ 5^2  \cdot 7 } \; , \\
q_{ss} & = & \frac{(5-7 \kappa^2)}{(25-21 \kappa^2)} \;, 
\eea 
where we used 
\bea 
& & \int^{\infty}_{-\infty} d w_{1} w_{1} 
\int^{\infty}_{-\infty} d w_{2} 
P( {\bf w} \vert \nu_{3}) \; d w_{3}
 = \nonumber \\
& & \mbox{}
\int^{\infty}_{-\infty} d w_{2} w_{2} 
\int^{\infty}_{-\infty} d w_{1} 
P( {\bf w} \vert \nu_{3}) \; d w_{3}
= \nonumber \\
& & \makebox[3.5cm]{} \; \;
 q_{ss} \overline{w}_{3} P( w_{3} \vert \nu_{3} ) d w_{3} \; .
\eea

\section{Calculations of Scale Functions}

\subsection{Ensemble Averaged Density of Peaks}  

The constraint of the peaks can be rewritten to 
\bea 
C(\mbox{\bf F}^{(10)} \vert  \mbox{\rm peaks} ) & = &
{\vert \lambda_1 \lambda_2 \lambda_3 \vert} \prod_{\alpha=1}^3  
\delta(F_{\alpha}) \Theta(\lambda_3) \nonumber \\ 
& = & \frac{(x-2z)[(x+z)^2-y^2]}{3^3} (\frac{\sigma_2}{\sigma_1})^3
\nonumber \\
& & \mbox{} \; \; \times 
\prod_{\alpha=1}^3 \delta( \nu_{\alpha} ) 
\Theta(x-2z) \;  . 
\eea
With the constraint,
in order to derive the ensemble-averaged number density of peaks,  
we will start from its probability weighted number density: 
\bea 
& & { n_{pk} (\nu, x, y, z; R) d\nu dx dy dz } 
\nonumber \\ 
& & \mbox{} \; \; = 6 \cdot P( \nu, \eta =\mbox{\bf 0}, x, y, z) 
  C(\mbox{\bf F}^{(10)} \vert {\rm peaks} ) d\nu dx dy dz  
\nonumber \\ 
& &  \mbox{} \; \; = 
\frac{5^{5/2} 3^{1/2}}{(2\pi)^3} ( \frac{\sigma_2}{\sigma_1})^3 
\frac{ \exp[-\frac{1}{2}\nu^2] \exp[-\frac{1}{2}Q(x,y,z)]}
{\sqrt{1-\gamma^2}} 
\nonumber \\
& & \makebox[1cm]{} \; \; \times
\psi_{pk}(x,y,z) \phi_{pk}(x,y,z) 
d\nu dx dy dz  \; , 
\eea
\bea
\psi_{pk}(x,y,z) & = & \frac{3^3}{2}\sigma_2^{-6} 
\lambda_1 \lambda_2 \lambda_3 
(\lambda_1-\lambda_2)(\lambda_2-\lambda_3)(\lambda_1-\lambda_3) \nonumber \\
& = & (x-2z)[(x+z)^2-(3y)^2]y(y^2-z^2)   \\
\phi_{pk}(x,y,z) & = &\left\{ \begin{array}{ll}  
           1 &\mbox{if} \; \frac{x}{4} \geq y \geq 0 \; \mbox{and} \; \\  
             &\makebox[1cm]{}  -y \geq z \geq y \nonumber \\
           1 &\mbox{if} \; \frac{x}{2} \geq y \geq \frac{x}{4} 
                        \; \mbox{and} \nonumber \\
             &\makebox[1cm]{}  3y-x \geq z \geq y \nonumber \\ 
           0 &\mbox{otherwise} \end{array} \right. \; , 
\eea 
where 
$ \phi (x,y,z) $ is the condition of $ \lambda_i > 0 $ and we had 
multiplied the probability expression by 6 to account for the ordering 
of the eigenvalues of $ \lambda_{i} $ .    

After the integration over $ y, z $, the ensemble-averaged density  
of the peaks is given as 
\bea  
& & < n_{pk}(\nu,x;R) > d\nu dx = 
\frac{ \exp[-\frac{1}{2}\nu^2]}{(2\pi)^{2} R_{\ast}^3 } d \nu
\nonumber \\
& & \mbox{} \; \; \times  
f_{pk}(x) \frac{\exp[ -(x-x_{\ast})^2/2(1-\gamma^2) ]} 
{\sqrt{2 \pi (1-\gamma^2)} } dx \; , 
\eea  
where $  R_{\ast}  =  \sqrt{3} \frac{\sigma_1}{\sigma_2} $ and 
\bea 
f_{pk}(x) & = & \frac{3^2 5^{5/2}}{\sqrt{2 \pi}}  
 [  \int^{x/4}_0 dy e^{-15 y^2 /2} 
   \int^{y}_{-y} \psi_{pk}(x,y,z) dz e^{-5 z^2 /2} \nonumber \\ 
 &   &  + \int^{x/2}_{x/4} dy e^{-15 y^2 /2} 
   \int^{y}_{3y-x} \psi_{pk}(x,y,z) dz e^{-5 z^2 /2}  
 ] \; , \nonumber \\
     & = & \frac{(x^3-3x)}{2} \{ 
{\rm Erf}[(5/2)^{1/2}x] + {\rm Erf}[(5/2)^{1/2}\frac{x}{2}] \} 
\nonumber \\
    & & \makebox[1cm]{}+ (\frac{2}{5\pi})^{1/2}[
    (\frac{31 x^2}{4} +\frac{8}{5})e^{-5x^2/8} 
\nonumber \\
    & & \makebox[2.5cm]{} \; 
 + (\frac{x^2}{2} -\frac{8}{5} ) e^{-5x^2/2} ] \; . 
\eea 

The density of the peaks of $ \nu $ at the scale $ R$ 
can be represented as 
\bea 
 < n_{pk}(\nu ;R) > d\nu = \frac{\exp (- \frac{1}{2}\nu^2)}
{(2\pi)^2 R_{\ast}^3} 
 G_{pk}( \gamma, x_{\ast}) d \nu \; , 
\eea 
where
\beq 
 G_{pk}( \gamma, x_{\ast}) =  \int^{\infty}_0 dx \; f_{pk}(x) 
\frac{\exp[ -(x-x_{\ast})^2/2(1-\gamma^2) ]} 
{\sqrt{2 \pi (1-\gamma^2)} } d \nu \; . 
\eeq
The function $ G_{pk}( \gamma, x_{\ast}) $ have a following 
fitting formula obtained by BBKS,    
\bea
G_{pk}( \gamma, x_{\ast})  & \simeq &  
\frac{ w^3 -3 \gamma^2 w + [ B w^2 + C_1 ] 
\exp [ -A w^2 ] }
{1 + C_2 \exp [ -C_3 w ] }\; ,  \\
A & = & \frac{5/2}{(9-5 \gamma^2)} \; , \\
B & = & \frac{432}{(10 \pi)^{1/2} (9-5\gamma^2)^{5/2} } \; , \\ 
C_1 & = & 1.84 + 1.13 (1-\gamma^2)^{5.72} \; , \\
C_2 & = & 8.91 + 1.27 \exp(6.51 \gamma^2) \; , \\ 
C_3 & = & 2.58 \exp(1.05 \gamma^2) \; .  
\eea 
We also have the averaged $ x $ ;
\bea 
  H_{pk}(\gamma,x_{\ast}) & = & \int_0^{\infty} x \; dx\; f_{pk}(x)
\nonumber \\
& & \makebox[1cm]{}
  \times \left\{ 
  \frac{ \exp[ -{(x-x_{\ast})^2}/{2(1-\gamma^2)} ]} 
       { \sqrt{2 \pi (1 -\gamma^2)} } \right\}   \nonumber  \\ 
  & = & < x >_{pk} G_{pk}(\gamma,x_{\ast}) \; . 
\eea 
We can easily see that the mean value $ < x >_{pk} $ is 
$ H_{pk}(\gamma,x_{\ast})/G_{pk}(\gamma,x_{\ast})$, which 
can be also represented as a approximated form as derived by BBKS :
\bea 
< x >_{pk} & = & \gamma \nu + \theta( \gamma, \gamma \nu ) \;, \\ 
\theta( \gamma, \gamma \nu ) & = & 
\frac{ \theta_1 + \theta_2  
\exp \left[ -\gamma/2(\gamma \nu/2)^2 \right] } 
{ \left[ \theta_1 + 0.45 + (\gamma \nu /2)^2 \right]^{1/2} 
+ \gamma \nu /2} \; , \\
\theta_1 & = & 3(1-\gamma^2) \; ,
\theta_2 = (1.216 - 0.9 \gamma^4) \; .
\eea

\subsection{Scale Functions of Peaks}  

The calculated ensemble-averaged densities is equal to
them of BBKS as 
\bea
< n_{pk} (\nu, x ;R) > d x d \nu  = {\cal N}_{pk} (\nu, x ;R)  d x d\nu \; .  
\eea 
With transformation of the density contrast $ \nu $ to the resolution
scale $ R $, we can obtain the scale function of peaks: 
\bea
& & N_{pk}(R,x, \delta_c) dR dx = {\cal N}_{pk} (\nu, x ;R) x
\left( \frac{\sigma_2}{\sigma_0} \right) R dR dx \;  \nonumber \\  
& & \makebox[1.5cm]{}  =  
\frac{ \exp[-\frac{1}{2}\nu^2]}{(2\pi)^{2} R_{\ast}^3 }
x \frac{\sigma_2(R)}{\sigma_0(R)}R dR \nonumber \\
& & \makebox[1cm]{} \times
f_{pk}(x) \frac{\exp[ -(x-x_{\ast})^2/2(1-\gamma^2) ]} 
{\sqrt{2 \pi (1-\gamma^2)} }  dx \; . 
\eea
This is the scale function of the peaks with the parameter of $ x$.  
With the integration of $ x $, 
the density function of peaks satisfying the condition of collapse
threshold $ \delta_c $ is given as  
\bea 
 N_{pk}(R,\delta_c) dR & = & \int_0^{\infty} x dx {\cal N}_{pk} (\nu_c, x, R) 
\frac{\sigma_2(R)}{\sigma_0(R)}R dR \nonumber  \\ 
& = & \frac{H_{pk}(\gamma,x_{\ast})}{(2\pi)^2 R_{\ast}^3} 
\exp[ -\frac{1}{2}\nu_c^2] \frac{\sigma_2(R)}{\sigma_0(R)}R dR \nonumber  \\ 
& = & {\cal N}_{pk} (\nu_c, R) <x>_{pk} 
\frac{\sigma_2(R)}{\sigma_0(R)}R dR \; . 
\eea 

\subsection{Scale Functions of The Nesting and Non-Nesting Peaks}

The above peak counting does not distinct nesting peaks and
non-nesting peaks.  In order to count the nesting peaks, we will introduce the 
joint scale function of the nesting peaks :  
\bea 
N^{nest}_{pk}(R_b,R_s ; \delta_c) d R_b d R_s & = &
N^{nest}_{pk}(R_s,| R_b ; \delta_c ) d R_s  \nonumber \\
& & \makebox[1cm]{} \times f (R_b,\delta_c ) d R_b \; ,
\eea   
where the fraction of the volume occupation with the objects
related to the filtering scale $ R_l$ is given as 
\bea 
f (R,\delta) d R  & = & \mbox{Vol}(R) \; N(R,\delta) d R
\; , \\ 
\mbox{Vol}(R) & = & \frac{M(R)}{\rho} \; ,
\eea
where $ N(R,\delta) d R $ is the scale function of non-nesting peaks.

In general, the conditional density of the peaks at $ \delta_s $ within 
the background at $ \delta_b $ on the larger scale $ R_b $ can be expressed 
as 
\bea 
& & N^{nest}_{pk}(R_s,\delta_s | R_b,\delta_b) d R_s = \nonumber \\
& & \mbox{} \; \;
{\cal N}_{pk}(R_s,\delta_s | R_b,\delta_b) 
\frac{\sigma_2(R_s)}{\sigma_0(R_s)}  
<\tilde{\underline{x}}> R_s \; d R_s \; , 
\eea 
in terms of the conditional density 
$ {\cal N}_{pk}(R_s,\delta_s | R_b,\delta_b) $ 
obtained by BBKS.  Here, the function $ <\tilde{\underline{x}}> = < x >_{pk} 
 ( \tilde{\gamma},\tilde{x}_{\ast} )  $ , where 
\bea 
\tilde{x}_{\ast} & = & \tilde{\gamma} \tilde{\nu} \;,  
\tilde{\gamma}= \gamma 
\sqrt{ 1 + \epsilon^2 \frac{(1-r_1)^2}{1-\epsilon^2} } \;, \\ 
\tilde{\nu} & = & \frac{\gamma}{\tilde{\gamma}}( \frac{1-r_1}{1-\epsilon^2}) 
\left[ \nu_s ( \frac{ 1- \epsilon^2 r_1}{1-r_1} ) -\epsilon \nu_b \right] 
\; . 
\eea 
The subscripts $ b $, $ s $ and $ h $ mean the values of the scales 
$ R_b $, $ R_s $, and $ R_h$ ($ R_b>R_h>R_s $ ).  According to BBKS, 
we introduced the following notations; 
\bea 
\epsilon = \frac{\sigma_{0h}^2}{\sigma_{0b} \sigma_{0s}} \; ,\\  
r_1=\frac{<k^2>_h}{<k^2>_s}=\frac{\sigma_{1h}^2 \sigma_{0s}^2} 
{\sigma_{0h}^2 \sigma_{1s}^2}=\frac{\sigma_{2h} \sigma_{0s}} 
{\sigma_{0h} \sigma_{2s} }  \;, \\
r_2=\frac{<k^3>_h}{<k^3>_s}=\frac{\sigma_{1h}^3 \sigma_{0s}^3} 
{\sigma_{0h}^3 \sigma_{1s}^3}=\frac{\sigma_{3h} \sigma_{0s}} 
{\sigma_{0h} \sigma_{3s} }  \;, \\
\sigma_{jh}= 4\pi \int dk k^{2j+2}\mid \delta_k \mid^2  
W(k;R_b) W(k;R_s) \; .
\eea
Since we use the Gaussian filter, we can rewrite as 
$ \sigma_{j \; h} = \sigma_j (R_h) $ with the rms average 
$  R_h= \left[ (R_b^2+R_s^2)/2 \right]^{1/2} $.  

This conditional density of peaks gives a form for  
$ N^{nest}_{pk}(R_s,| R_b ; \delta_c ) d R_s $ 
at the the limit of $ \delta_b \rightarrow \delta_s = \delta_s $.
In this paper,
we use this simple form for the conditional density as 
\bea 
& & N^{nest}_{pk}(R_s,| R_b ; \delta_c ) d R_s = \nonumber \\
& & \mbox{} \; \; {\cal N}_{pk}(R_s,\delta_c | R_b,\delta_c) 
\frac{\sigma_2(R_s)}{\sigma_0(R_s)}  
<\tilde{\underline{x}}> R_s \; d R_s \; . 
\eea   

The number density of the non-nesting peaks $ N( R, \delta_c ) $
can be obtained by subtracting the density of the pairs
of nesting peaks 
$ N^{nest}(R, \delta_c) $  from that of all peaks $ N_{pk}( R, \delta_c) $, 
which includes the nesting peaks $ N^{nest}(R, \delta_c) $   
at $ \delta_c $ 
on the scale $ R $.  Then, the notations 
to the corrected scale function is 
\bea 
N (R,\delta_c) dR & = &  N_{pk}(R,\delta_c)dR  -N^{nest}(R,\delta_c) dR \; . 
\eea

The scale function of the nesting peaks can be calculated from
the integration of the joint scale function of
the nesting peaks $ N^{nest} (R,
R' ; \delta_c) d R_c d \delta_c $,
\bea 
N^{nest} (R, \delta_c) d R & = & 
\int^{\infty}_{R} d R'
N^{nest} (R', R ; \delta_c) d R_c
\; \nonumber \\
& = & 
\int^{\infty}_{R} d R'
f (R', \delta_c ) \nonumber \\
& & \makebox[1cm]{} \times N^{nest}_{pk}( R | R' ; \delta_c ) d R \; .
\eea
This is a Volterra type integral equation, which can be solved
with the iteration process in exact mean as shown in Manrique $\&$
Salvador-Sole (1995).  The number density 
of the non-nesting peaks $ N (R,\delta_c) $ can be also obtained
with the same iteration process.
The number density of the nesting peaks can be negligible to
that of all peaks, as shown in Fig. 2. 

\subsection{Ensemble Averaged Density of Sloping Saddles} 

The constraint of the sloping saddles are rewritten to 
\bea 
& & C( \mbox{\bf F}^{(20)} \vert \: \mbox{s. saddles} ) d R = 
\nonumber \\
& & \makebox[0.5cm]{} \; \; 
\frac{1}{3} \frac{\sigma_2 \sigma_3}{\sigma_1^2} \vert (x+3y+z)(x-2z)
w_3 (w_1 +w_2 +w_3) \vert
\nonumber \\
& & \makebox[0.5cm]{} \; \; \times 
\delta(z-3y+x) \delta^{(3)} ( \eta ) 
\left( \frac{\sigma_3}{\sigma_1} \right) R dR \; . 
\eea 
With the constraint,
the probability weighted number density of the sloping saddles
is represented as  
\bea 
& & {\bf n}_{ss}(\nu,x,y,z,{\bf w} ;R )d \nu dx dy dz
d^3{\bf w} d R \nonumber \\
& & \makebox[0cm]{} = 6 \cdot 
P({\bf w} \vert \nu_3 =0 ) d^3 {\bf w}   
    P(x, y, z \vert \nu)dx dy dz P( \nu, \eta =0 ) d\nu \nonumber \\
& & \makebox[1.5cm]{} \times 
    C(\mbox{\bf F}^{(20)} \vert \mbox{s. saddles} ) d R 
   \nonumber \\ 
& & \makebox[0cm]{} = \frac{ 3^{5/2} \cdot 5^{5/2} }{(2 \pi)^3} 
(\frac{ \sigma_2 \sigma_3}{\sigma_1^2}) R \psi_{ss}(x,y,z) 
\phi_{ss}(x,y,z) \nonumber \\
& & \makebox[0.5cm]{} \times \frac{\exp[-\frac{1}{2}\nu^2] 
\exp[-\frac{1}{2}Q(x,y,z)]}{ \sqrt{ 1-\gamma^2 }} 
d\nu dx dy dz \nonumber \\
& & \makebox[0.5cm]{} \times
\vert w_3 (w_1 +w_2 +w_3) \vert
P ({\bf w} \vert \nu_3=0 ) d^3 {\bf w} \nonumber \\
& & \makebox[0.5cm]{} \times
\left( \frac{\sigma_3}{\sigma_1} \right) R dR \; ,
\eea
\bea 
\psi_{ss} (x,y,z) & = & y(y^2 -z^2)(x+3y+z) \nonumber \\
                  & & \makebox[1cm]{} \; \times (x-2z) \delta(z-3y+x) \; , \\ 
\phi_{ss} (x,y,z) & = & \phi_{pk} (x,y,z)\; , 
\eea
where the factor 6 is needed to account for the ordering of $
\lambda_i $ the same as that of the peaks.
With the integration over $ y, z $, 
we can obtain the 
density of the sloping saddles similar to that of the peaks;  
\bea 
& & < {\bf n}_{ss}(\nu, x, {\bf w}; R ) > d \nu dx d^3{\bf w} d R
 \nonumber \\
& & \makebox[0.5cm]{} =
\frac{\exp[-\frac{1}{2}\nu^2 ]d\nu }
{(2 \pi)^{2} R_{\ast}^3 } \nonumber \\
& & \makebox[1cm]{} \times  
f_{ss}(\kappa, x)  
\frac{ \exp[ -(x-x_\ast)^2/ 2 
(1-\gamma^2)] } { \sqrt{2 \pi (1-\gamma^2)}} dx  \nonumber \\
& & \makebox[1cm]{} \times
\vert w_3 (w_1 +w_2 +w_3) \vert P({\bf w} \vert \nu_3=0 ) d^3 {\bf w}
\nonumber \\
& & \makebox[1cm]{} \times
\left( \frac{\sigma_3}{\sigma_1} \right) R dR \;, 
\eea 
where we used the relation 
$ ( \sigma_2 \sigma_3 /\sigma_1^2  )^{1/3} 
 = \sqrt{3}/ \kappa^{1/3} /R_\ast  $,  and 
\bea 
 f_{ss}(\kappa, x) & = & A_{ss}(\kappa) \int^{x/2}_{x/4} 
dy {\cal Y}(y ; x) \nonumber \\
& = & A_{ss}(\kappa) e^{-5x^2/8} 
[\frac{4}{5^3 \cdot 3}(1-e^{-15x^2/8}) \nonumber \\
& & \makebox[2cm]{} \; \; 
+ \frac{1}{5} x^2(\frac{3 x^2}{2^5} -\frac{1}{10}) ] \; , 
\eea 
\bea 
 A_{ss}(\kappa) & = &  
\frac{ 3^{4} \cdot 5^{5/2} }{ \sqrt{2\pi} \kappa} 
\; , \\
{\cal Y}(y ; x) & = & e^{-15y^2/2} \int^y_{3y-x} \psi_{ss}(x,y,z) 
e^{-5z^2/2} dz  \nonumber \\
& = & 18  y^2 (x-2y)^2 (4y-x) \nonumber \\
& & \makebox[1cm]{} \times 
\exp [ -\frac{15 y^2}{2} - \frac{5 (3y-x)^2}{2} ] \; . 
\eea 

The density averaged with {\bf
w} is given  
\bea 
& & < \mbox{\bf n}_{ss}(\nu, x ;R ) > dR d x d \nu 
=    
\frac{\exp[-\frac{1}{2}(\nu^2)] d\nu}
{(2 \pi)^{2} R_{\ast}^3 } \nonumber \\
& &  \makebox[1cm]{} 
\times f_{ss}(x)  
\frac{ \exp[ \frac{-(x-x_\ast)^2}{2 
(1-\gamma^2)} ] } { \sqrt{2 \pi (1-\gamma^2)}} dx \nonumber \\
& &  \makebox[1.5cm]{} \times  
<w^2>_{ss}(1+q_{ss}) \frac{\sigma_3(R)}{\sigma_1(R)}R dR \; . 
\eea 
With the integration over $ x $, the averaged density is 
\bea 
& & < \mbox{\bf n}_{ss}( \nu ; R) > dR d \nu \nonumber \\
& & \makebox[1cm]{} =
\frac{G_{ss} (\gamma,x_{\ast})e^{-\frac{1}{2}\nu^2} }
{(2 \pi)^2 R_{\ast}^3} 
d \nu  \nonumber \\
& & \makebox[2cm]{} \times 
< w_3^2>_{ss}(1 + q_{ss})
\frac{\sigma_3(R)}{\sigma_1(R)}R dR \; , 
\eea 
where 
\bea 
G_{ss}(\gamma,\kappa,x_{\ast}) & = & \int^{\infty}_{0} f_{ss}(x)   
\frac{ \exp [- \frac{(x-x_\ast)^2}{2 (1-\gamma^2)} ] }
{\sqrt{2 \pi (1-\gamma^2)}}  dx \; . 
\eea 
The exact result of the integration is obtained as an analytical form 
\bea 
& & G_{ss}(\gamma,\kappa,x_{\ast}) =  
\frac{ A_{ss}(\kappa) }{\sqrt{2\pi(1-\gamma^2)}} 
\lbrack \nonumber \\
& & \; \; 
\frac{ 2 (1-\gamma^2) x_{\ast} }{ 5 (9-5\gamma^2)^3 }
\lbrace{     
      \frac{ 12 x_{\ast}^2 }{ (9 - 5\gamma^2) } 
     +\frac{ ( 39 - 55 \gamma^2 ) }{5} 
\rbrace} 
       e^{ - \frac{ x_{\ast}^2 }{ 2( 1-\gamma^2) } } 
\nonumber \\ 
& & \mbox{}  +  \sqrt{ \frac{2\pi( 1 - \gamma^2)}{( 9 - 5\gamma^2)^9} } 
\lbrace 
         \frac{24 x_{\ast}^4}{5}
        +\frac{4(27 - 35\gamma^2)( 9 - 5\gamma^2) x_{\ast}^2 }{25}
\nonumber \\
& & \makebox[2.5cm]{} +
        \frac{(783-1230 \gamma^2 +575 \gamma^4)( 9 - 5\gamma^2)^2}{750} 
\rbrace
\nonumber \\
&  & \makebox[1cm]{}\times 
\{
1 + {\rm Erf}\left[ x_{\ast} \sqrt{ \frac{2}{(9 - 5\gamma^2)(1-\gamma^2)} } 
             \right]
\} 
   e^{ - \frac{5 x_{\ast}^2}{2( 9 - 5\gamma^2)} }
\nonumber \\
&  & - 
\frac{2}{375} 
\sqrt{ \frac{2\pi(1 -\gamma^2)}{(6-5\gamma^2)} } 
\{ 
1 + {\rm Erf}\left[ \frac{x_{\ast}}{\sqrt{2( 6-5\gamma^2)(1-\gamma^2)}}
             \right]
\} \nonumber \\
&  & \makebox[1cm]{} \times 
e^{- \frac{5 x_{\ast}^2}{2(6-5\gamma^2)} }
\rbrack \; . 
\eea

The mean value of $ x $ at the sloping saddles is 
\bea 
< x >_{ss} & = & 
\frac{H_{ss}(\gamma,\kappa,x_{\ast})}{G_{ss}(\gamma,\kappa,x_{\ast})}\; ,\\  
H_{ss}(\gamma,\kappa,x_{\ast}) & = & \int^{\infty}_{0} x \; f_{ss}(x)   
\frac{ \exp[ -\frac{(x-x_\ast)^2}{2 (1-\gamma^2)} ] }
{\sqrt{2 \pi (1-\gamma^2)}}  dx \; , 
\eea 
where we can also obtain the analytical result of the integration :
\bea 
& & H_{ss}(\gamma,\kappa,x_{\ast})  =  
\frac{ A_{ss}(\kappa)}{ \sqrt{2\pi(1-\gamma^2)} } 
\lbrack  \nonumber \\
&  & \makebox[0.5cm]{}   
\frac{4(1-\gamma^2)}{(9-5\gamma^2)^3}  
\lbrace \frac{24 x_{\ast}^4 }{ 5( 9 - 5\gamma^2)^2 }  
 +\frac{ 2 ( 99 - 115 \gamma^2 )x_{\ast}^2 }{ 25 ( 9 -5\gamma^2) }
\nonumber \\
& & \makebox[3cm]{}
 + \frac{ 9( 1 - \gamma^2 )^3 }{ 6 - 5\gamma^2 }  
\rbrace
      e^{ - \frac{ x_{\ast}^2 }{ 2( 1-\gamma^2) } } 
\nonumber \\ 
&  & \makebox[0.5cm]{} +  
\frac{2}{5} 
\sqrt{ \frac{2\pi( 1 - \gamma^2)}{( 9 - 5\gamma^2)^7} } 
\lbrace  \frac{48 x_{\ast}^5}{( 9 - 5\gamma^2)^2 }
         +\frac{8( 57 - 65\gamma^2)x_{\ast}^3}{5( 9 - 5\gamma^2)}
\nonumber \\ 
&  & \makebox[2cm]{}   
         +\frac{( 2403 - 4950\gamma^2 + 2675 \gamma^4 )x_{\ast}}{75} 
\rbrace 
\nonumber \\
&  & \makebox[1cm]{} \times 
\lbrace 
1 + {\rm Erf}\left[ x_{\ast}\sqrt{ \frac{2}{(9 - 5\gamma^2)(1-\gamma^2)} } 
             \right]
\rbrace
    x_{\ast} \; 
    e^{ -\frac{5 x_{\ast}^2}{2( 9 - 5\gamma^2)} }
\nonumber \\
&  & - 
\frac{2 }{375} 
\sqrt{ \frac{2\pi (1 -\gamma^2)}{(6-5\gamma^2)^3} } 
\lbrace{ 
1 + {\rm Erf}\left[ x_{\ast} \frac{1}{ \sqrt{2( 6-5\gamma^2)(1-\gamma^2)} } 
\right]
\rbrace} \nonumber \\
&  & \makebox[2cm]{} \times 
x_{\ast} \; e^{ - \frac{5 x_{\ast}^2}{2(6-5\gamma^2)} \; \;  } 
\rbrack
\eea 

\section{Some Formula in the cases for Power Law Spectrum} 

When the power spectrum can be represented as a simple power law as 
\beq 
\mid \delta_k \mid^2  = A k^n \; ,  
\eeq 
we can obtain the useful relations for the calculations as  
\bea 
\sigma_j(R) & = &  A^{1/2} \sqrt{2 \pi} R^{-(\frac{n+3}{2}+j)} 
\Gamma(\frac{n+3}{2}+j)^{1/2} \; ,  \\
\frac{\sigma_i(R)^2}{\sigma_j(R)^2} &=&
      \frac{(2i+n+1)!!}{2^{i-j}(2j+n+1)!!} R^{-2(i-j)} \;  , \\ 
R_{\ast}&=&(\frac{6}{n+5})^{1/2} R\; , \\
\gamma^2 &=&( \frac{\sigma_1^2}{\sigma_0 \sigma_2} )^2 = 
\frac{(n+3)}{(n+5)} \; ,  \\ 
\kappa^2 &=& ( \frac{\sigma_2^2}{\sigma_1 \sigma_3})^2 
= (\frac{n+5}{n+7}) \;, \\
\epsilon &=&( \frac{R_s R_b}{R_h^2} )^{(n+3)/2} \; ,\\
r_1&=&(\frac{R_s}{R_h})^2 \;, \\
r_2&=&(\frac{R_s}{R_h})^3  \; . 
\eea
%\\
%\bea
%\frac{\sigma_1^2(R)}{\sigma_0^2(R)} &=& \frac{(n+3)}{2} R^{-2} \;  , 
%\frac{\sigma_2^2(R)}{\sigma_0^2(R)}=\frac{(n+5)(n+3)}{4} R^{-4} \;  , \\
%\frac{\sigma_2^2(R)}{\sigma_1^2(R)} &=& \frac{(n+5)}{2} R^{-2} \;  , 
%\frac{\sigma_3^2(R)}{\sigma_1^2(R)}=\frac{(n+7)(n+5)}{4} R^{-4} \;  , \\
%\frac{\sigma_3^2(R)}{\sigma_2^2(R)} &=& \frac{(n+7)}{2} R^{-2} \;  , \\
%\eea
\end{appendix}

\end{document}